\documentclass[acmsmall]{acmart}

\setcopyright{cc}
\setcctype{by}
\acmJournal{PACMHCI}
\acmYear{2026} \acmVolume{10} \acmNumber{6} \acmArticle{CSCW049}
\acmMonth{10} \acmDOI{10.1145/3816897}

\received{13 May 2025}
\received[revised]{13 January 2026}
\received[accepted]{17 March 2026}

\AtBeginDocument{%
  }

\newcommand{\ailego}{\mbox{AI LEGO}}
\newcommand{\ailegolite}{\mbox{\textsc{AI LEGO Lite}}}
\newcommand{\ailegofull}{\mbox{\textsc{AI LEGO Full}}}

\newcommand{\dvlifecycleblocks}{\textsl{Lifecycle Blocks}}
\newcommand{\dveightstageworksheet}{\textsl{Eight-Stage Worksheet}}
\newcommand{\dveightstagechecklist}{\textsl{Eight-Stage Checklist}}
\newcommand{\dvstagecenteredeval}{\textsl{Stage-centered Evaluation}}
\newcommand{\dvpersonacenteredeval}{\textsl{Persona-centered Evaluation}}
\usepackage{enumitem}
\usepackage{rotating}
\usepackage{framed}
\usepackage{multirow}

\begin{document}

\title[Scaffolding Cross-Functional Collaboration in Industrial Responsible AI Practices during Early Design Stages]{\ailego: Scaffolding Cross-Functional Collaboration in Industrial Responsible AI Practices during Early Design Stages}


\author{Muzhe Wu}
\orcid{0009-0004-6842-0180}
\authornote{Both authors contributed equally to this research.}
\affiliation{%
  \institution{Carnegie Mellon University}
  \city{Pittsburgh}
  \state{PA}
  \country{USA}
}
\affiliation{%
  \institution{University of Michigan}
  \city{Ann Arbor}
  \state{MI}
  \country{USA}
}
\email{muzhewu@gmail.com}

\author{Yanzhi Zhao}
\orcid{0009-0002-2180-0798}
\authornotemark[1]
\affiliation{%
  \institution{Carnegie Mellon University}
  \city{Pittsburgh}
  \state{PA}
  \country{USA}
}
\email{yanzhiz@alumni.cmu.edu}

\author{Shuyi Han}
\orcid{0009-0007-2607-9342}
\affiliation{%
  \department{Northwestern University}
  \institution{Northwestern University}
  \city{Evanston}
  \state{IL}
  \country{USA}
}
\email{shuyihan2026@u.northwestern.edu}

\author{Michael Xieyang Liu}
\orcid{0000-0002-8246-8736}
\affiliation{%
  \institution{Carnegie Mellon University}
  \city{Pittsburgh}
  \state{PA}
  \country{USA}
}
\email{lxieyang.cmu@gmail.com}

\author{Hong Shen}
\orcid{0000-0002-5364-3718}
\affiliation{%
  \institution{Carnegie Mellon University}
  \city{Pittsburgh}
  \state{PA}
  \country{USA}
}
\email{hongs@cs.cmu.edu}
\renewcommand{\shortauthors}{Wu et al.}

\begin{abstract}
Responsible AI (RAI) efforts increasingly emphasize the importance of addressing potential harms early in the AI development lifecycle through social-technical lenses. However, in cross-functional industry teams, this work is often stalled by a persistent coordination challenge: how technical roles hand off technical intent, how teams establish shared structures for collaboration, and how non-technical roles are supported in systematically evaluating harms. Through literature review and a semi-structured interview study with 8 practitioners, we unpack how this challenge manifests---technical design choices are rarely handed off in ways that support meaningful engagement by non-technical roles; collaborative workflows lack shared, visual structures to support mutual understanding; and non-technical practitioners are left without scaffolds for systematic harm evaluation. Existing tools like JIRA or Google Docs, while useful for product tracking, are ill-suited for supporting joint harm identification across roles, often requiring significant extra effort to align understanding. To address this, we developed \ailego, a web-based prototype that operationalizes the boundary object theory to support cross-functional AI practitioners in effectively facilitating knowledge handoff and identifying harmful design choices in the early design stages. Technical roles use interactive blocks to draft development plans, while non-technical roles engage with those blocks through stage-specific checklists and LLM-driven persona simulations to surface potential harms. In a study with 18 cross-functional practitioners, \ailego\ increased the volume and likelihood of harms identified compared to baseline worksheets. Participants found that its modular structure and persona prompts made harm identification more accessible, fostering clearer and more collaborative RAI practices in early design.
\end{abstract}


\begin{teaserfigure}
  \centering
  \includegraphics[width=\textwidth]{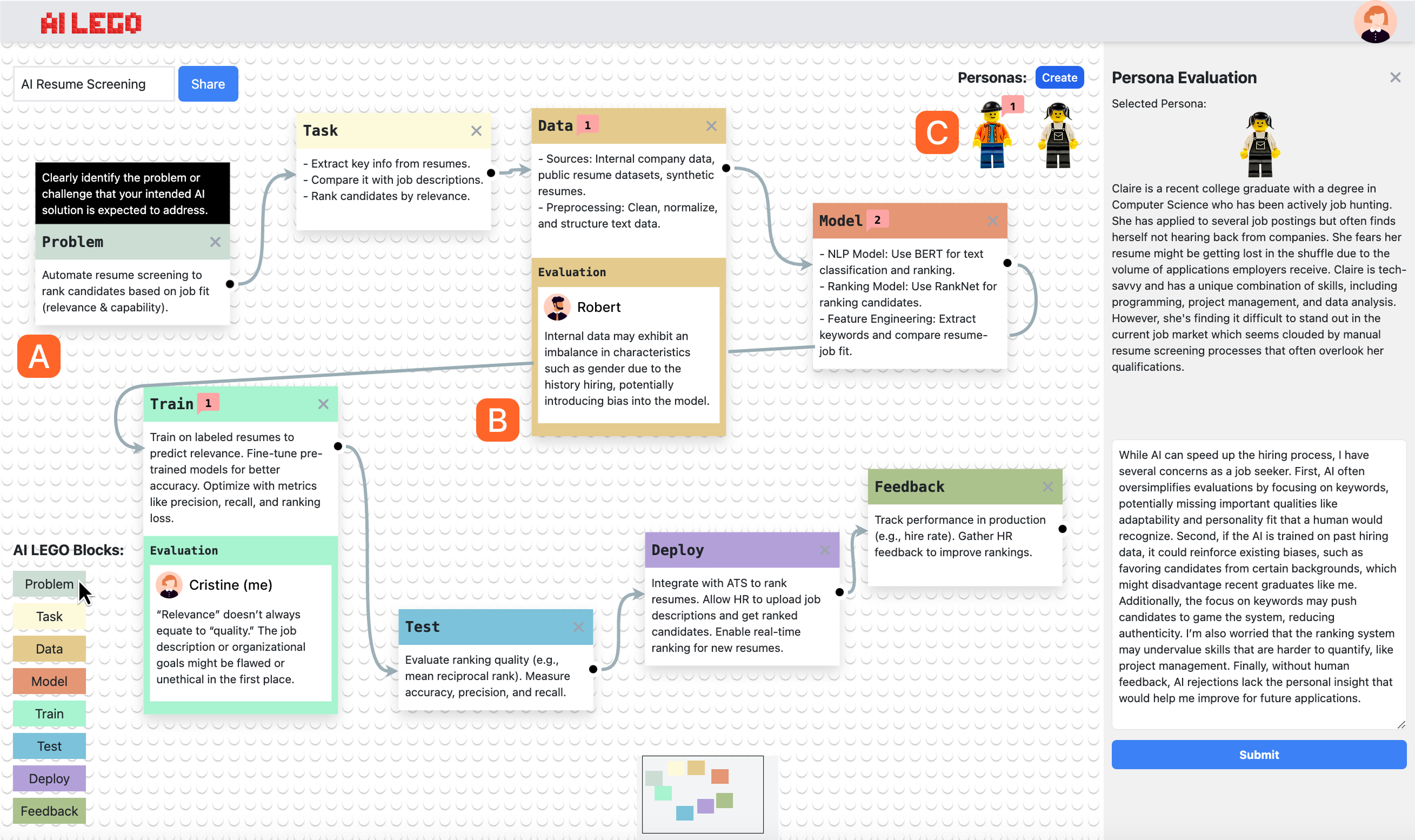}
  \vspace{-7mm}
  \Description{Screenshot of the AI LEGO web interface illustrating the three-stage workflow. Technical AI developers first create a development plan using color-coded lifecycle blocks connected into a workflow. Non-technical collaborators then attach stage-specific evaluations to individual lifecycle stages. They can also use persona-centered evaluation to generate stakeholder personas and simulate their perspectives to help identify additional potential harms. The example shown uses an AI resume screening application.}
  \caption{\ailego\ is a web tool that scaffolds cross-functional collaboration in industrial AI development planning and proactive harm identification. With \ailego, technical AI developers can sketch out a development plan using (A) \dvlifecycleblocks, each corresponding to a specific AI development stage with a tailored prompt. They hand off the plan to non-technical/user-facing roles, who conduct (B) \dvstagecenteredeval\ to systematically identify and address potentially harmful design choices at each stage. Lastly, (C) \dvpersonacenteredeval\ feature helps generate stakeholder personas to surface edge cases by simulating diverse perspectives. Minifigure images © seewhatmitchsee / iStock by Getty Images. Used under license.}
  \Description{}
  \label{fig:teaser}
\end{teaserfigure}

\maketitle
\section{Introduction}

Artificial Intelligence (AI)-driven systems, powered by Machine Learning (ML) techniques, have exercised power over many aspects of our everyday lives, from deciding which social media posts we see, the quality of healthcare we receive, to the ways we are represented to potential employers~\cite{noble2018algorithms,eubanks2018automating,gillespie2014relevance,narayanan2024ai}.
The rise of Generative AI has further amplified this influence. Yet they have also raised pressing concerns, as public outcries and scholarly work have increasingly documented the ways these systems can, whether inadvertently or intentionally, perpetuate or even amplify existing societal biases or create new harmful impacts ~\cite{barocas2016big,buolamwini2018gender,sweeney2013discrimination,shen2021everyday}.

Alongside the establishment of high-level governmental policies and frameworks (e.g., NIST's AI Risk Management Framework in the United States~\cite{ai2023artificial}), researchers in Responsible AI (RAI) and Human-Computer Interaction (HCI) have developed various technical tools and interactive systems to support model evaluation, auditing, and debiasing~\cite{bellamy2019ai,bird2020fairlearn}.
Despite their significant contribution, this line of work has often been criticized for (1) failing to surface certain ``blind spots'', often due to the lack of diversity in \textit{technical} AI teams ~\cite{holstein2019improving,deng2023understanding}; and (2) being conducted only after the system is built or deployed, when significant harms had already occurred ~\cite{veale2018fairness,deng2022exploring}.

Recent CSCW and HCI research has increasingly emphasized the socio-technical nature of AI systems and the importance of involving non-technical/user-facing roles, such as designers, product/project managers (PMs), and policy stakeholders, in RAI work~\cite{chancellor2023toward,green2020algorithmic,liao2024ux}. In industry, such work is often carried out by \textit{cross-functional AI teams}, where participants bring diverse forms of expertise~\cite{deng2023investigating}. Prior studies show that involving non-technical roles in these teams can help surface problematic design assumptions in the early development process~\cite{wang2023designing,wong2023seeing}.

However, existing work also indicates that early-stage collaboration in cross-functional AI teams is not simply a matter of participation, but of coordination~\cite{schmidt1992taking,ackerman2000intellectual, passi2018trust,zhang2020data}. When AI designs are still evolving and only partially specified, teams must collaboratively reason about technical intent, constraints, and potential downstream impacts across roles with asymmetric expertise. While prior research has identified communication challenges in such settings~\cite{piorkowski2021ai,almahmoud2021teams,nahar2022collaboration,deng2023investigating}, how collaborative work can be structured to support this form of early-stage RAI coordination, before implementation decisions become fixed, remains insufficiently specified~\cite{wong2023seeing,madaio2024tinker,rakova2021responsible}.

Through a formative interview study with eight cross-functional AI practitioners, we examine how early-stage RAI collaboration breaks down in practice. When AI designs are still evolving and only partially specified, teams must collaboratively reason about technical intent, design trade-offs, and potential downstream harms across roles with asymmetric expertise. Our study surfaced three recurring challenges that shape this coordination work. First, technical design rationales and constraints are often difficult to externalize in ways that support meaningful engagement by non-technical collaborators, leading to breakdowns in knowledge handoff~\cite{liu2021reuse,deng2023investigating}. Second, collaborative practices frequently rely on ad hoc documentation and communication, leaving teams without shared structure to support discussion, reflection, and iteration~\cite{piorkowski2021ai,rakova2021responsible}. Third, even when design information is available, non-technical practitioners often lack scaffolds for systematically interpreting design choices and evaluating potential downstream harms~\cite{wang2023designing,wong2023seeing}. Together, these challenges point to a broader coordination gap in early-stage cross-functional RAI collaboration.

Building on CSCW theories of articulation work~\cite{schmidt1992taking} and boundary objects~\cite{star1989institutional}, we view these challenges as intertwined aspects of a coordination problem rather than isolated breakdowns. From this perspective, harm identification is not a separate or downstream activity, but depends on whether technical intent can be articulated, shared, and interrogated across roles. Supporting early-stage RAI collaboration therefore calls for tools that integrate articulation, shared collaborative structure, and role-appropriate scaffolds for interpretation and evaluation within a single workflow.

This interpretation leads us to ask the following question: \textbf{How can boundary objects be designed to support early-stage cross-functional RAI collaboration across roles with asymmetric technical expertise?} We derived three design goals to address the challenges: (1) supporting articulation and handoff of technical design intent, (2) providing shared and flexible structures to organize cross-functional collaboration, and (3) scaffolding systematic harm evaluation by non-technical/user-facing roles. We instantiated these goals in \ailego\ (Fig. \ref{fig:teaser}), an interactive web tool that operationalizes the boundary object theory~\cite{star1989institutional} to help cross-functional AI practitioners
more effectively communicate high-level design ideas and identify potential problematic design choices that may lead to downstream harms in the early design stage.

We further conducted an evaluation user study with 18 cross-functional AI practitioners grouped into 6 teams to investigate (1) the effectiveness of our tool in helping cross-functional practitioners with RAI work in the early design stage; (2) the value of our tool over existing RAI practices for cross-functional AI practitioners.  The results show that \ailego\ helped AI practitioners identify an average of 195\% more problematic design choices that would lead to downstream harms with a higher likelihood of occurring (1.25-point increase in median on a 4-point scale), compared to using Google Docs and Harm Modeling framework~\cite{microsoft2022harms} baseline. Participants also reported that \ailego\ was effective in bridging gaps in knowledge handoff, providing a flexible collaborative structure, and supporting multi-perspective harm evaluation.

The contributions of this work include:

\begin{itemize}[topsep=0pt,leftmargin=0.3in]
\item \textbf{Design goals} from literature review and a formative semi-structured interview study to tackle the challenges of scaffolding early-stage cross-functional RAI collaborations, including knowledge handoff from technical roles to non-technical/user-facing roles, lack of shared collaborative structure, and limited scaffolds for harm evaluation.

\item \textbf{\ailego, an interactive web tool} to support cross-functional AI practitioners in effectively facilitating knowledge handoff
and identifying harmful design choices in the early design stage, when issues are easier to fix without causing actual harm. 
\item \textbf{Empirical findings} from an evaluation study on using \ailego\ during early-stage cross-functional RAI collaboration practices.
\ailego\ helped participants identify more high-quality potential harms in AI designs and is more useful compared to existing resources.
\item \textbf{Design implications} on supporting effective cross-functional collaboration in conducting RAI work, especially in the early-stage AI design.

\end{itemize}

\section{Related Work}
We situate our work at the intersection of (1) RAI tools and methods that support anticipating and addressing harms, and (2) CSCW research on coordination across heterogeneous roles through shared representations, articulation work, and boundary objects. We first review RAI toolkits and early-stage design methods, then examine cross-functional RAI collaboration as a coordination problem, and finally discuss harm identification and stakeholder reasoning approaches that inform our system design.

\subsection{Developing techniques, toolkits and systems for Responsible AI in industry}
Over the years, the RAI community in HCI/CSCW has made significant contributions in developing techniques, toolkits, and systems to support industry practitioners in conducting RAI-related work across a wide range of AI systems~\cite{chen2018my,holstein2019improving,angell2018themis,madaio2020co,wang2024farsight}. 
Many early tools focused on supporting technically trained practitioners in evaluating trained models, with a particular emphasis on fairness, bias, and performance trade-offs~\cite{wexler2019if,bird2020fairlearn,bellamy2019ai}. While these tools have been influential in operationalizing RAI principles, prior work has noted that they often surface issues late in the development lifecycle, when core design choices, such as data sources, proxies, or task definitions, are already difficult to revise~\cite{veale2018fairness,deng2022exploring}.

Beyond model-centric tools, a parallel line of work has developed ethics toolkits and design methods aimed at supporting broader, cross-functional reflection on ethical and social impacts. These include card-based and workshop-oriented approaches such as Envisioning Cards~\cite{friedman2012envisioning}, Value Cards~\cite{shen2021value}, and practitioner-facing toolkits such as IDEO’s AI Ethics Cards~\cite{ideo2019aiethicscards}. Such methods explicitly seek to engage designers, developers, and business stakeholders in collaborative discussions of ethical considerations early in the design process.
However, prior work also cautions that ethics toolkits can implicitly encode assumptions about who performs ethics work and how ethical concerns translate into accountable, coordinated action, potentially leaving gaps around workflow integration, ownership, and power dynamics~\cite{wong2023seeing}.

Recent work has begun exploring earlier-stage RAI interventions that move upstream in the design process. For example, Farsight~\cite{wang2024farsight} supports harm anticipation during early prototyping of LLM-powered applications, and Lam et al.~\cite{lam2023model} introduce frameworks for iteratively authoring approximations of model decision logic. While these approaches push RAI considerations earlier, they are typically designed for technical practitioners and do not explicitly structure handoff or evaluation across cross-functional roles. Our work builds on these efforts by focusing on how early-stage design artifacts can be structured to support collaborative reasoning and evaluation across roles with asymmetric expertise.

\subsection{Cross-functional collaboration in Responsible AI}
CSCW has long examined how collaborative work is coordinated across heterogeneous roles, emphasizing articulation work, shared information spaces, and the persistent negotiation required to align specialized forms of expertise~\cite{suchman1987plans,strauss1988articulation,schmidt1992taking,ackerman2000intellectual}. In organizational settings, cross-functional collaboration often depends on shared representations that allow participants to coordinate despite differences in goals, vocabularies, and evidentiary standards.

Empirical studies of industry AI development show that such coordination is particularly challenging in AI work, where design decisions span technical, organizational, and social considerations. Passi and Jackson describe how corporate data science projects rely on continual ``translation'' work to make technical decisions intelligible and accountable to non-technical stakeholders~\cite{passi2018trust}. Zhang et al.~\cite{zhang2020data} and Piorkowski et al.~\cite{piorkowski2021ai} further document how AI teams depend on a patchwork of tools, documents, and meetings to coordinate across roles, often adapting general-purpose artifacts (e.g., tickets, shared documents) to communicate design intent and constraints. However, these adaptations are fragile: they frequently leave key assumptions implicit, privilege technical framings, or require substantial interpretive labor from non-technical collaborators.

Prior literature highlights that cross-functional collaboration around fairness and ethics introduces additional layers of complexity. Studies show that fairness concepts are interpreted differently across roles, that evidence takes multiple forms (e.g., metrics, anecdotes, lived experience), and that responsibility for ethical decisions is often diffuse or contested~\cite{deng2023investigating,rakova2021responsible,liao2024ux}. Wong et al.~\cite{wong2023seeing} further argue that many RAI tools implicitly assume how ethical work should proceed, without adequately accounting for how teams actually coordinate this work over time. Together, this literature suggests that participation alone is insufficient: effective early-stage RAI work depends on how collaboration is structured to support articulation, interpretation, and negotiation across roles with asymmetric expertise.

A core CSCW concept for understanding such coordination is \textit{boundary objects}---shared yet flexible artifacts that enable collaboration across epistemic and professional boundaries~\cite{star1989institutional,carlile2002pragmatic}. CSCW work has examined how boundary objects support coordination in settings such as healthcare and collaborative planning, where artifacts must remain interpretable and actionable for participants with different expertise~\cite{reddy2001coordinating,schmidt1997maps}. Importantly, boundary objects are not ``solutions'' on their own: they often require ongoing articulation work to keep representations meaningful and aligned with situated practice~\cite{schmidt1992taking,ackerman2000intellectual}. 
In RAI contexts, boundary objects have been used to support shared reflection and accountability.
For example, Madaio et al.~\cite{madaio2024tinker} have proposed to contextualize fairness checklists into cross-functional AI teams' collaborative workflows and balance individual ownership of fairness decisions with the oversight of a designated approver.
However, existing work rarely specifies how these artifacts are actually worked through over time—how they are handed off, interpreted, and evaluated as AI design decisions evolve across development stages and roles.

We developed \ailego\ to further explore and operationalize boundary objects as part of a structured, staged workflow that explicitly supports knowledge handoff~\cite{liu2021reuse,clases2002steps}, collaborative evaluation, and iteration in early-stage RAI work. Rather than treating coordination as an implicit outcome of discussion, we examine how coordination can be designed for through shared artifacts that evolve as design decisions are examined across roles.

\subsection{Identifying harmful problems in AI systems, products and services}

Identifying harmful or unethical issues in technology has been a long-standing concern in HCI ~\cite{friedman1996value,auger2013speculative,video-tape}. As AI systems have become more integrated into society, addressing these concerns has taken on even greater importance. Over the years, a wide range of design methods and toolkits have been developed to support designers, developers, and data scientists in recognizing and anticipating potential harms in AI systems~\cite{shen2021value,friedman2012envisioning,madaio2020co,ballard2019judgment,doteveryone2020agile}. 
For example, 
Raji et al.~\cite{raji2020closing} define a meta-level, end-to-end framework for AI system audit, with the goal to ensure the accountability of the system prior to deployment.
On the other hand, Qiang et al.~\cite{qiang2024no} conducted a series of comparative studies and found that different AI ethics frameworks serve complementary purposes and are most effective when applied in combination, depending on the context.

One prominent approach to supporting RAI evaluation is the use of personas and stakeholder representations~\cite{cooper2003face,chen2024empathy,shen2021value}. Personas provide concrete stand-ins for affected groups and serve as boundary objects that help teams reason about social impacts and lived experiences~\cite{cooper2003face}.
Prior work, such as Value Cards toolkit~\cite{shen2021value}, incorporates Persona Cards alongside Model Cards and Checklist Cards, demonstrating effectiveness in facilitating discussions on the social impacts of ML models. 
With the rise of generative AI models like LLMs, this approach can be further augmented to accelerate and inspire human evaluation in RAI work. For example, Buçinca et al. ~\cite{buccinca2023aha} introduced a general framework that assists AI practitioners and decision-makers in anticipating potential harms of AI systems, using LLMs to generate descriptions of potential harms for various stakeholders. In a similar vein, Farsight ~\cite{wang2024farsight} leverages LLMs to help practitioners build LLM-based applications to ideate different use cases, identify stakeholders, and consider potential harms.

We build on and extend this body of work by integrating it into the context of cross-functional collaboration at the early design stage. \ailego\ introduces two scaffolding techniques specifically designed to help non-AI developers more effectively identify problematic AI design choices early in the process. First, we developed a stage-based checklist to support harm anticipation, which covers key design choices across the entire AI development lifecycle for its effectiveness~\cite{kuo2023understanding}. Second, building on prior work~\cite{buccinca2023aha,wang2024farsight}, we incorporated LLMs to assist non-AI developers in brainstorming potentially impacted stakeholders and gaining a deeper understanding of their lived experiences.
By treating these scaffolds as boundary objects, integrated into a shared workflow that structures how AI design decisions are examined across roles, we explore how they can support interpretive alignment while also surfacing tensions between machine-generated suggestions and human ethical reasoning.
\section{Formative Study and Design Goals}
To better understand (1) the challenges that interdisciplinary teams face when identifying potential harms in the early stage of AI development, and (2) how features in cross-functional collaboration tools or frameworks can be developed to support this process, we conducted a formative semi-structure interview study\footnote{The formative study along with the evaluation study were approved by our institution's IRB (STUDY2023\_00000435).} as detailed below.

\subsection{Semi-Structured Interview}
We recruited 8 industrial AI practitioners (4 female, 4 male) via social media and mailing lists.
These participants held diverse roles across three major categories (4 AI developers\footnote{In this work, we broadly use the term ``AI developers'' to refer to practitioners who implement, train, or maintain AI/ML systems, including roles such as AI/ML Engineer, Data Scientist, and AI/ML Research Scientist.}, 2 PMs, and 2 UI/UX designers)~\cite{constantinides2024rai} and work at companies of varying sizes\footnote{Four from companies with 5,000--24,999 employees, and one each from companies with 1,000--4,999, 250--999, 50--249, and fewer than 50 employees.
}.
Participants were screened and selected based on their experience in cross-functional AI development.
They were not expected to have prior experience in RAI, as informed by prior work~\cite{madaio2024tinker}, reflecting the typical demographics of AI practitioners aiming for broader understanding.
All participants were based in the United States due to logistical constraints and to ensure alignment with regional industry practices.
Each participant was compensated with a \$35 USD Amazon gift card.
Each study session lasted approximately 60 minutes and focused on identifying specific hurdles a participant has faced in interdisciplinary teamwork and exploring how different tool designs can enable efficient early-stage collaboration to mitigate problematic AI design. We began by introducing the goal of our study, then asked participants to describe their current workflows and challenges encountered in cross-functional collaboration between technical and non-technical/user-facing roles. This helped us understand their existing practices and pain points.
We then presented five preliminary design artifacts (see Appendix~\ref{sec:codesign-ui} for the designs and their rationales), created by the researchers prior to the session, to the participants.
These artifacts included two approaches for surfacing AI development plans: plain description and stage-based scaffolding inspired by AI storyboarding~\cite{cramer2019challenges}, and three UIs for brainstorming potential harms: a survey adapted from Value Cards~\cite{shen2021value}, stakeholder table, and commenting.
Participants were asked for their opinions on whether and how these designs could support their current workflows, their perceived usability, and suggestions for improvement (see Appendix~\ref{sec:formative-interview-protocol} for the full semi-structured interview protocol).

\subsection{Findings}
We analyzed participants' (P1--8) notes with affinity diagrams ~\cite{lucero2015using}.
Two authors independently reviewed all collected materials and generated initial observations by clustering related comments, sketches, and annotations. Through an iterative process, the authors collaboratively grouped these observations into three higher-level clusters:

\subsubsection{Articulating and Handing Off Technical Intent} 
Participants consistently described early-stage cross-functional collaboration as challenging not simply due to a lack of communication, but due to difficulties in articulating technical intent in forms that non-technical collaborators could meaningfully engage with.
While participants noted that communication flows and patterns within AI teams can vary depending on factors such as team size (P2), organizational or business type (P4, P8), and individual differences (P3), most of them ($n=8$) agreed that initiating an AI development plan is best led by technical roles or practitioners with strong technical expertise to ensure feasibility (e.g., P3 noted, ``\textit{My PM often overestimates what the model can achieve.}'')
This arrangement, however, would often result in breakdowns during handoff to non-technical or user-facing roles ($n=5$). 

Several participants ($n=3$) emphasized that existing documentation and planning artifacts were either overly technical or fragmented, making it difficult for non-technical collaborators to understand how design decisions were made or where potential risks might arise. For example, P6 noted that ``\textit{explicating data construction in plain language would be a prerequisite for success},'' reflecting a broader need for accessible representations of technical decisions. In the absence of such representations, some non-technical participants reported a strong dependence on technical roles for interpretation and decision-making ($n=2$), limiting their autonomy in contributing to early design discussions.

Participants also highlighted how commonly used coordination tools—such as JIRA~\cite{jira}, Asana~\cite{asana}, and Google Docs~\cite{google_docs}—were poorly suited for this form of early-stage RAI work. These tools were described as either too rigid to support fast iteration or too generic to accommodate discussions of values and harms. As P8 remarked, ``\textit{JIRA takes a lot of time to set up and isn’t suitable for fast turnarounds}.'' As a result, teams often relied on ad hoc explanations or synchronous meetings to establish shared understanding, which several participants ($n=4$) found inefficient or unnecessary.

\subsubsection{Establishing Shared Structure for Early-stage AI Planning} 
Participants expressed a strong preference for structured yet flexible representations to support early-stage AI development planning across roles. Comparing the two artifacts, all participants ($n=8$) preferred stage-based scaffolding, consistently viewing it as an effective structure for coordinating work and establishing a shared frame of reference. Participants believed that stage-based scaffolding allowed technical developers to clearly outline tasks and dependencies within AI development, while non-technical members benefited from its clear and modularized visuals. According to P2, ``stage-based representation standardizes the language of AI development planning'', thus being more easily graspable for both technical and non-technical/user-facing roles. The stage-level breakdown also ``facilitates frequent iteration'' (P1), which is especially helpful in early-stage AI practices.
 
\subsubsection{Scaffolding Interpretive Work in Harm Identification}
Participants ($n=4$) appreciated the straightforwardness and in-depth investigation of the survey adapted from Value Card~\cite{shen2021value}, albeit the amount of reading appeared discouraging (P1, P2).
In contrast, the stakeholder table had a clear information visualization ($n=4$), allowing for ``both horizontal and vertical comparisons'' (P1) across stages and stakeholders for the comprehensiveness of evaluation. 
Commenting was highlighted for its ability to foster open-ended discussions and effectively link identified problems to specific design choices ($n=3$), facilitating faster iteration. Ideal harm identification scaffolding should integrate these characteristics.

\subsection{Design goals}
Combining the insights from the literature review and the semi-structured interview study, we establish the following design goals:

\begin{itemize}[leftmargin=0.25in]

\item[\textbf{G1}] \textbf{Support articulation and handoff of technical intent.}  
The system should enable technical roles to externalize early AI development decisions at an appropriate level of abstraction, allowing non-technical and user-facing collaborators to meaningfully engage with and evaluate.

\item[\textbf{G2}] \textbf{Provide shared, flexible structure for collaborative planning.}  
The system should offer visual, modular representations that organize early AI development work while remaining adaptable to different project contexts and collaboration styles.

\item[\textbf{G3}] \textbf{Scaffold interpretive evaluation by non-technical and user-facing roles.}  
The system should integrate structured yet open-ended evaluation supports that help non-technical practitioners reason about potential harms across development stages and stakeholder perspectives, while preserving space for exploratory and situated judgment.

\end{itemize}
\section{AI LEGO}
Guided by the design goals, we developed AI LEGO, an interactive web tool that helps cross-functional AI practitioners collaboratively identify harmful design choices in the early stages of AI design.
We begin by showcasing a sample scenario of a cross-functional team collaborating on an RAI project to demonstrate the AI LEGO workflow.

\subsection{Example Usage Scenario}
A startup movie platform is developing a personalized recommendation model. John, an AI developer, is tasked with creating the development plan but struggles to present it to the rest of the team. Emma, the product manager, is worried about potential biases, like recommendations influenced by gender or age, associated with problematic design decisions.

They decide to use AI LEGO for support. John starts outlining the key AI development stages in the project using AI LEGO's \dvlifecycleblocks\ (Fig.~\ref{fig:teaser}A). The embedded \dveightstageworksheet\ prompts him to input details for each stage, creating a clear visual map of the development plan.
When John shares the plan with Emma, the clear visual framework and tailored prompts immediately provide Emma with an overarching understanding of the development plan. Emma then uses AI LEGO's \dvstagecenteredeval\ (Fig.~\ref{fig:teaser}B) to review key stages like ``Dataset Construction'' and ``Model Definition.'' The tool's \dveightstagechecklist\ guides her in identifying risks, such as sensitive data like gender potentially causing biased recommendations. She suggests anonymizing this data but raises concerns about its impact on accuracy.
To explore further, Emma uses the \dvpersonacenteredeval\ (Fig.~\ref{fig:teaser}C) to assess stakeholder impact. She creates a few personas, including a teenage movie enthusiast, which helps her realize the need for age-appropriate recommendations. This prompts a reassessment of how sensitive data is handled to balance fairness and accuracy.

With these insights, John revises the development plan, adjusting the dataset and model design to address ethical concerns. After a few iterations, John and Emma finalized a well-rounded plan ready for actual development.

\begin{table}[t]
\renewcommand{\arraystretch}{1.0}
\centering
\Description{Three-column table listing the eight AI development stages supported by AI LEGO. For each stage, the table pairs a worksheet prompt used by technical practitioners to describe design decisions with a checklist prompt used by non-technical practitioners to identify potential ethical concerns. Stages include problem formulation, task definition, dataset construction, model definition, training, testing, deployment, and feedback.}
\caption{The Eight-Stage Worksheet and Checklists offer prompts tailored to different stages of AI development, guiding AI developers in drafting the development plan and assisting non-technical/user-facing roles in identifying problematic design choices.}
\label{tab:eight-stage-prompts}
\resizebox{1\textwidth}{!}{%
\begin{tabular}{p{34mm}p{76mm}p{65mm}}
\toprule
\textbf{Stage Name}           & \textbf{Worksheet Prompt} & \textbf{Checklist Prompt} \\
\noalign{\global\arrayrulewidth=0.8pt}\midrule
\noalign{\global\arrayrulewidth=0.4pt}
\textbf{Problem Formulation}  & Clearly identify the problem or challenge that your intended AI solution is expected to address. & Is the problem or challenge itself ethical? Can the intended AI solution provide a viable solution to the identified problem? \\
\midrule
\textbf{Task Definition}      & Detail the specific task(s) that your intended AI solution is designed to perform to solve the previously identified problem. & What are the boundaries and limitations of what the AI solution is expected to achieve? \\
\midrule
\textbf{Dataset Construction} & Describe where and how the data for training your intended AI solution is collected and prepared, and explain the data preprocessing steps and any feature engineering technologies that are applied to the raw data. If the intended AI solution is a generative AI, describe the fine-tuning dataset you're using. & Are there any potential biases, privacy concerns, and other ethical considerations in data handling? \\
\midrule
\textbf{Model Definition}     & Describe the AI model architecture, the proxies selected, and the algorithms used, explaining their roles and functionalities. If the intended AI solution is a generative AI, describe the base model you're using, the prompting techniques, and the fine-tuning methods. & Is the choice of models, proxies, and algorithms appropriate for the task identified?
If the intended AI solution is a generative AI, is the choice of the base model, the prompting techniques, and the fine-tuning methods appropriate for the task identified? \\
\midrule
\textbf{Training}             & Describe how the AI solution is trained using the curated data, and clarify the process of how it improves its performance. & Can you think of ways the training process might go wrong? If so, how? \\
\midrule
\textbf{Testing}              & Explain how the AI solution is evaluated and highlight the key performance metrics used to assess the AI. & What would it mean for the AI to be successful? Are the performance metrics sufficient for evaluating the success of AI? \\
\midrule
\textbf{Feedback}             & Outline how feedback is collected, specifying who the feedback is gathered from and how often the feedback is collected. & Does the deployment of the AI fit with the real-world practical use environments? \\
\bottomrule
\end{tabular}
}
\end{table}

\subsection{Eight-stage Worksheet \& Checklist}
\dveightstageworksheet\ and \textit{Checklist} (Table~\ref{tab:eight-stage-prompts}) are both lists of prompt questions designed to, respectively, elicit high-level descriptions of key design choices in AI development from technical roles and support non-technical/user-facing roles in potential harm investigation.
Based on findings from prior research~\cite{schmidt1992taking,schmidt1996coordination} and our formative study findings on the effectiveness of ``staging'', we developed prompts that address eight stages of the AI development lifecycle: problem formulation, task definition, dataset construction, model definition, training, testing, and feedback.
These stages are adapted from Cramer et al.’s seven-stage workflow~\cite{cramer2019challenges}, with the addition of an explicit problem formulation stage. Prior work~\cite{kuo2023understanding} has similarly incorporated this stage and highlights its critical role in shaping downstream technical decisions in RAI practice~\cite{passi2019problem}.
The curation of prompt questions took an iterative process, ensuring that they not only capture socio-technical aspects, but also scaffold effective articulation of technical intent or non-technical/user-facing concerns and meaningful reflection around crucial aspects of each stage.

\subsection{User Interface}
\ailego\ further integrates \dveightstageworksheet\ and \textit{Checklist} into a visual, interactive interface consisting of the following components (see Fig.~\ref{fig:teaser}; UI transitions detailed in Fig.~\ref{fig:ui-components} in Appendix~\ref{sec:ui-components}):

\textbf{Lifecycle Blocks.}
The \dvlifecycleblocks\ are stage-based visual building blocks used to map out the development plan of an intended AI system (G2).
Technical AI developers can create blocks in the central canvas by clicking on the stage button in the bottom left corner (Fig.~\ref{fig:teaser}A). When they click on the block's textbox, a corresponding prompt from the \dveightstageworksheet\ will appear above, guiding them to fill in the relevant descriptions. Once the descriptions are completed, they can link the blocks and reposition them using drag-and-drop functionality to establish a flow or structure. The canvas expands infinitely in all directions, allowing developers to add as many blocks and branches as needed for maximum flexibility.

\textbf{Stage-centered Evaluation.}
\dvstagecenteredeval\ enables non-technical/user-facing roles to examine the individual stages in a drafted plan to identify potential problematic design choices (G3).
Users start by selecting a block and pressing the ``Stage'' button in the bottom-right corner. This generates a sidebar displaying the corresponding prompt from the \dveightstagechecklist\ and allowing users to describe the problematic designs. The completed evaluation is automatically saved, attached to the corresponding stage block, and shared within the team (Fig.~\ref{fig:teaser}B).

\textbf{Persona-centered Evaluation.}
To support perspective-taking and reflective brainstorming during early-stage harm evaluation, \ailego\ introduces \dvpersonacenteredeval, a feature that uses personas as scaffolds for exploring potential stakeholder viewpoints. During the evaluation process, non-technical roles can envision any stakeholders who might be affected by the intended AI system and create personas with specified characteristics such as personal and professional backgrounds using text descriptions. To balance user control and diverse representation, \dvpersonacenteredeval\ supports a combination of manual input and LLM-generated examples. These personas are saved and visualized in the top-right corner as LEGO Minifigure icons (Fig.~\ref{fig:teaser}C).
After identifying the impacted stakeholders, \dvpersonacenteredeval\ can further inspire evaluators by simulating personas' reactions to the proposed AI development plan. Non-technical roles can choose a saved persona and inquire about their subjective feelings. To enable more open-ended discussions, we prompt LLMs to simulate the concerns stakeholders might articulate based on their backgrounds, rather than directly identifying harms~\cite{buccinca2023aha}. These simulated reactions are saved and displayed when hovering over the persona icon.

\subsection{Implementation Details}
The \ailego\ web application is built in HTML, JavaScript, and CSS with the React library.
Its node-based interface is enabled with vanilla React plus react-draggable\footnote{https://www.npmjs.com/package/react-draggable} package.
While the frontend is hosted on Vercel, the backend uses Google Firebase through API calls for user authentication and data persistence (per state change). The \dvpersonacenteredeval\ functionality leverages the OpenAI GPT-4 model with zero-shot prompting techniques (see Appendix~\ref{sec:prompts}).
\section{Evaluation User Study}
We conducted a user study to evaluate the effectiveness of \ailego\ in helping cross-functional AI practitioners communicate and identify problematic design choices of AI systems at an early stage.
To approximate real-world cross-functional collaborative settings while maintaining a controlled environment for an initial empirical understanding of our tool, we organized participants into teams and divided the study sessions into two phases.
We adopted a within-subject design and created two variants of our tool, \ailegolite\ and \ailegofull.
Both variants allow practitioners to draft AI development plans with \dvlifecycleblocks. Yet \ailegolite\ only supports \dvstagecenteredeval\ while \ailegofull\ supports both \textit{Stage-centered} and \dvpersonacenteredeval.

Our study was guided by the following Research Questions: \textit{\textbf{RQ1}}: How effective are \ailegofull\ and \ailegolite\ in assisting cross-functional AI practitioners?; \textit{\textbf{RQ2}}: Do \ailegofull\ and \ailegolite\ offer value over existing RAI practices for cross-functional AI practitioners?

\subsection{Participants \& Team Formation}
We recruited 18 industry AI practitioners (6 female, 12 male) through social media posting and snowball sampling.  None of these participants were involved in our previous study. Similarly, all participants were physically located in the United States and were required to have prior industry experience working on AI-infused products in cross-functional teams, but were not expected to have worked on RAI. In total, we received 101 voluntary responses.
Based on pre-screening survey responses about their roles in past AI projects, we selected 6 AI developers, 6 PMs, and 6 UI/UX designers\footnote{A total of 5 participants worked at companies with 25,000 or more employees. Others came from companies with 5,000--24,999 employees (3), 1,000--4,999 (2), 250--999 (1), 50--249 (1), and fewer than 50 (4).}.
Participants were randomly assigned to one of six teams before the study. Each team included one technical role (AI developer) and two non-technical/user-facing roles (one PM and one UI/UX designer), who participated in different tasks during the study (see Section~\ref{sec:study-phase}).
Participants within each team were unidentified/unacquainted with each other due to the recruitment procedure. Future work could investigate the dynamics with acquaintances.
Each participant received a \$35 USD Amazon gift card after the study session.

\subsection{Study Design}
\label{sec:study-phase}

We designed a within-subject two-phase study to evaluate the effectiveness of \ailego\ in cross-functional RAI practices:

\begin{figure}[tp]
\centering
\includegraphics[width=0.75\textwidth]{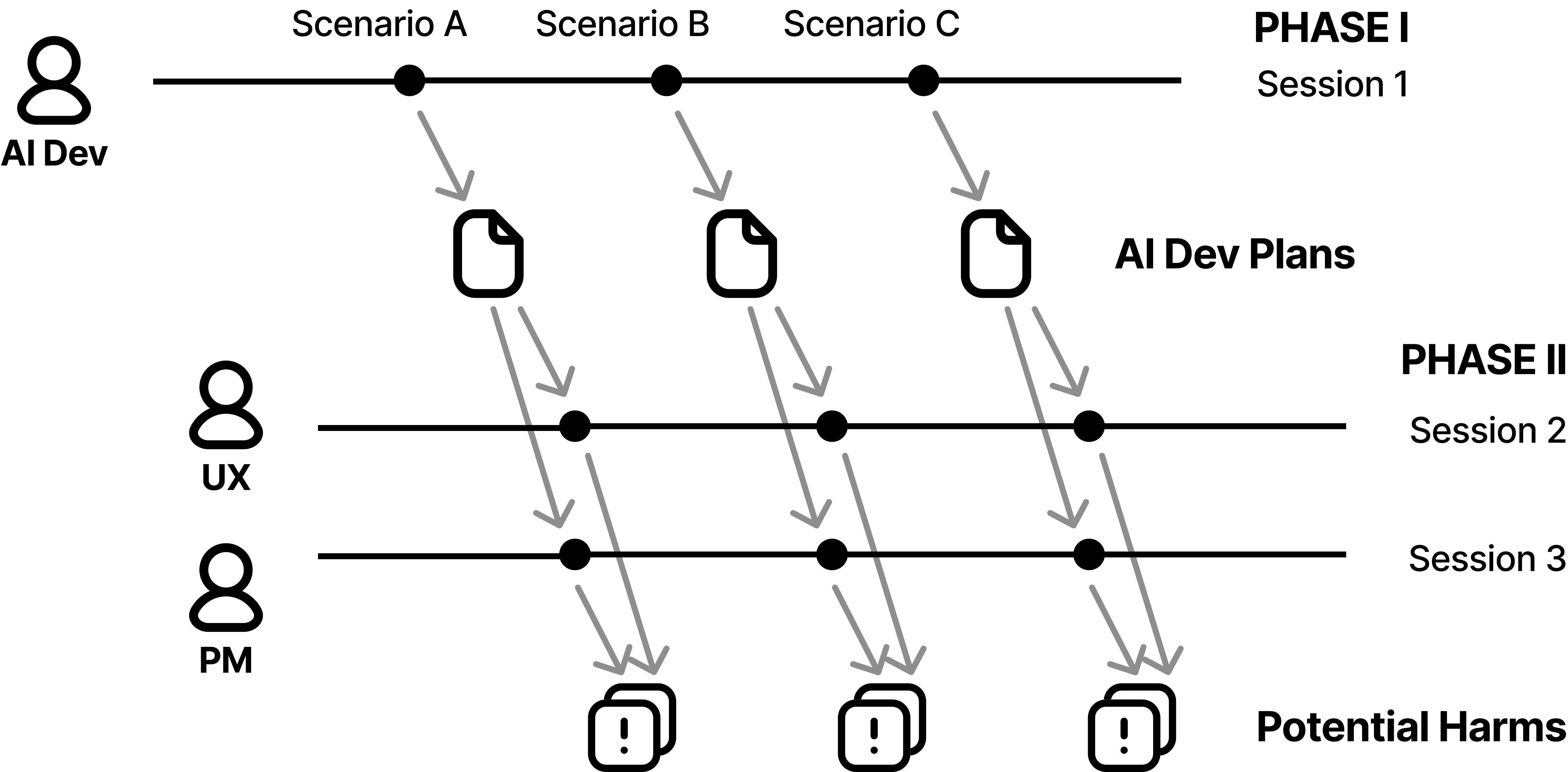}
\Description{Workflow diagram of the two-phase evaluation study. In Phase I, an AI developer creates development plans for three AI scenarios in three different experimental conditions. In Phase II, a UX designer and product manager independently review each plan to identify potential harms.}
\caption{Two-phase team-based evaluation study task flow.}
  \label{fig:study-design}
\end{figure}

\subsubsection{Two-phase Design}
The overall task flow within a team (3 sessions) is shown in Fig.~\ref{fig:study-design}.
In \textbf{Phase I}, the technical AI developer initiated the process by drafting AI development plans.
In three blocks, they were tasked with creating three AI development plans, each focusing on a specific scenario (see Section~\ref{sec:study-scenario}) and leveraging a different tool corresponding to the conditions (see Section~\ref{sec:study-condition}).
In \textbf{Phase II}, the PM and UX designer separately reviewed the plans created by the AI developer within their team in the prior session and were tasked with reviewing the plan to identify potential downstream harms that arose from problematic design choices.
They were presented with the plans created by the AI developer and the scenario description, applying tools corresponding to the conditions.\looseness=-1

\subsubsection{Conditions}
\label{sec:study-condition}
Each participant experienced all three different conditions/tools: Baseline (Google Docs), \ailegolite, and \ailegofull. 
Google Docs is one of the most popular online collaboration and documentation tools commonly used in industry settings. It facilitates a high degree of freedom in text input and styling.
We further complement Harm Modeling~\cite{microsoft2022harms}, a popular risk identification framework (as an editable table) in the doc. Together, they provide a reasonably strong baseline for scaffolding RAI practices including planning and identifying potential risks.
\ailegolite\ and \ailegofull\ are identical in Phase I for scaffolding AI development plan drafting.
In Phase II, \ailegolite\ supports only \dvstagecenteredeval, whereas \ailegofull\ supports both \dvstagecenteredeval\ and \dvpersonacenteredeval.
This contrast was intentionally designed to examine the incremental contribution of \dvpersonacenteredeval\ to the harm identification process when layered on top of the structured, stage-centered evaluation workflow.
The order of conditions was fixed for participants within each team and was counterbalanced across teams through Latin squares~\cite{bradley1958complete}.

\subsubsection{AI System Development Scenarios}
\label{sec:study-scenario}
To mimic typical industrial early-stage AI design processes, we created five scenarios (see Appendix~\ref{sec:scenarios}) where AI systems are to be developed to address real-world problems. These scenarios were sourced and adapted from the AI Incident Database~\cite{mcgregor2021preventing}.
All sampled incidents had significant media coverage across diverse AI application domains.
Each scenario includes information about the background, goals, requirements, evaluation metrics, key features, deployment details, and target users---representing a typical starting point for early-stage AI design. We removed all identifiable entity names from the materials to avoid biases. We ensured that all team members were unfamiliar with the original incident through a pre-screening survey.
Each team was assigned three out of the five scenarios in a fixed order.
Across all teams, each scenario was assigned to each condition/tool at least twice. The detailed scenarios and conditions assignment is shown in Fig.~\ref{fig:scenario-condition-order} in Appendix~\ref{sec:assignment}.

\subsubsection{Session Procedure}
Each session lasted approximately 60 minutes and was conducted one-on-one via remote video conferencing.
At the start, participants were introduced to the goal of the study and provided their consent for recording their screen activity and audio.
Participants then progressed through the three blocks, where they performed tasks according to their roles in designated conditions.
Each block took approximately 15 minutes and ended when participants reported reaching thought exhaustion.
At the end of a block, they evaluated the tool's usability by filling the System Usability Scale~\cite{brooke1996sus}. For Phase I, SUS scores for \ailegolite\ were recorded only for the first intervention. Participants were offered a voluntary 2-minute break between conditions. The session concluded with a 10-minute semi-structured interview (see Appendix~\ref{sec:eval-interview-protocol} for the protocol), where they shared their preferences over the conditions, reasons behind their choices, and additional feedback.
\subsection{Data Analysis}
Data analysis aims to assess the effectiveness and usability of \ailegofull\ and  \ailegolite\ compared to Google Docs in early-stage RAI practices.
We analyzed both quantitative (ratings of identified harm, SUS scores, Preference) and qualitative data (semi-structured interview notes) with methodologies described as follows.

\subsubsection{Quantitative Analysis}
Following ~\cite{wang2024farsight}, we analyzed the count, severity, and likelihood of identified harms to evaluate the effectiveness of \ailego\ (RQ1).
After collecting the identified potential harms through system logging and recording, two authors independently rated the severity (``How severe is the harm's impact in this scenario?'') and likelihood (``How likely is this harm to occur in this scenario?'') of all the harms using a 4-point scale.
Both authors were graduate students with academic \& industry experiences in RAI (unlike many of the participants).
They were blinded to the conditions under which harms were identified but had access to the scenarios for contextual understanding.
The initial ratings showed a fair to moderate agreement with weighted (quadratic) Cohen's Kappa scores of $0.51$ for severity and $0.64$ for likelihood.
After the initial ratings, the authors resolved discrepancies through discussion and developed a grading rubric based on Sociotechnical Harms Taxonomy~\cite{shelby2023sociotechnical}.
The final grading rubric with examples are provided in Table~\ref{tab:rubric} in Appendix~\ref{sec:rubric}.

The interval data (count of identified harms) were aggregated at participant level using the mean and analyzed using a repeated measures ANOVA, followed by paired t-tests with Holm correction for pairwise comparisons. This approach was appropriate as the Shapiro--Wilk test indicated no significant deviation from normality ({\small $p>.05$}). Mauchly’s test indicated a violation of the sphericity assumption ({\small $p = .02$}) and thus Greenhouse--Geisser correction was applied. The ordinal data (severity, likelihood of identified harms, SUS scores, average preference rankings) was aggregated at participant level using the median and analyzed with non-parametric Friedman tests, followed by Wilcoxon signed-rank tests with Holm correction for post-hoc pairwise comparisons. For SUS scores and preference rankings in Phase I, the Wilcoxon signed-rank test was applied instead as there were only two conditions.

\subsubsection{Qualitative Analysis}
We transcribed the semi-structured interview recordings and applied inductive thematic analysis~\cite{goddard2011semantic}. Two authors independently conducted an initial open coding pass over the transcripts, focusing on participants’ descriptions of their experiences with the tools, perceived utility, limitations, and impacts on cross-functional RAI practices (RQ2).
The authors then met to compare and reconcile their codes, iteratively grouping related codes into higher-level themes. These themes were refined through repeated examination of the transcripts to ensure they captured recurring patterns across participants while preserving different perspectives.

\subsubsection{Power and Sensitivity Analysis}
To assess whether our study design was capable of detecting meaningful within-subject effects, we conducted a sensitivity analysis using G*Power for a one-group repeated-measures ANOVA with three conditions ({\small $\alpha = .05$}). We focus this analysis on Phase II, which constitutes the primary confirmatory evaluation of system effects, while Phase I is intended to support exploratory design insights. Assuming a moderate correlation among repeated measures ({\small $r = .5$}) and a conservative Greenhouse--Geisser correction ({\small $\epsilon = .65$}, consistent with our observed data), the Phase II design ({\small $n = 12$}) was sensitive to effects of approximately Cohen’s {\small $f \ge 0.46$} with 80\% power. This indicates that the study was sufficiently sensitive to detect moderate-to-large condition effects, though smaller effects may not have been reliably detected. Limitations, including demographic composition and implications for generalizability, are discussed in Section~\ref{sec:limitations}.
\section{Findings}
\subsection{Effectiveness of \ailego\ in Assisting Cross-Functional AI Practitioners (RQ1)}
\label{sec:results-quant}
\begin{figure}[t]
    \centering
    \includegraphics[width=\linewidth]{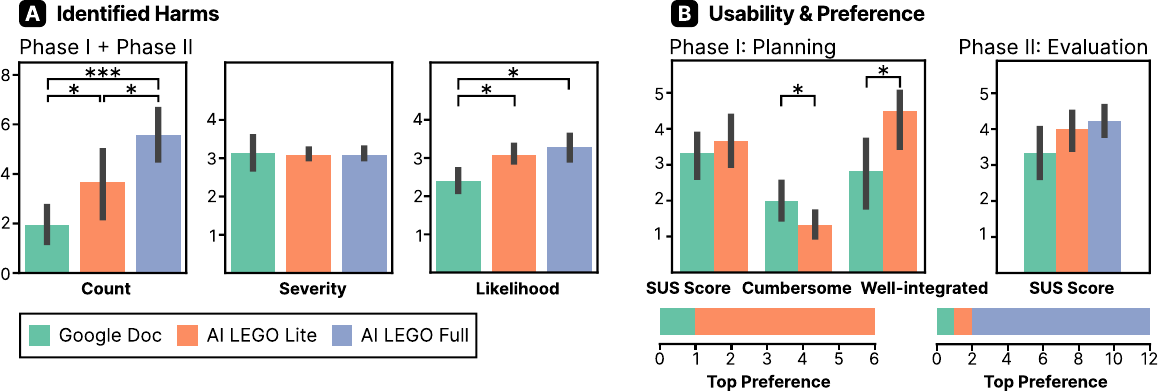}
    \Description{Bar charts comparing Google Docs, AI LEGO Lite, and AI LEGO Full. AI LEGO Full results in the highest number of identified harms and the highest likelihood ratings while maintaining similar usability scores. Participants rank AI LEGO as their preferred tool during both planning and evaluation phases and considered it less cumbersome and more well-integrated than Google Docs for planning. Error bars represent 95\% confidence intervals and significant pairwise differences are indicated.}
    \caption{Effect of conditions on (A) Count, Severity, and Likelihood of identified harms and (B) Usability and Preference. \ailegofull\ allowed AI teams to identify more potential harms with a higher likelihood; it was more preferred and considered well-integrated, especially in the planning phase. Significance levels: {\small *** $p < .001$, ** $p < .01$, * $p < .05$}. Error bars indicate 95\% CI.}
    \label{fig:results}
\end{figure}

Quantitative results demonstrated varied effectiveness of \ailegofull\ and \ailegolite\ in helping cross-functional teams proactively identify problematic design choices to prevent potential downstream harms (Fig.~\ref{fig:results}A). Participants also rated \ailegolite\ and \ailegofull\ to be more usable and preferable compared to Google Doc (Fig.~\ref{fig:results}B).

\subsubsection{\ailego\ enables the cross-functional team to uncover more problematic design choices.}
The repeated measures ANOVA test revealed a significant main effect of conditions ({\small $F(2, 22)=16.26, p_{GG}<.001$}) on the number of problematic design choices identified. Post-hoc Holm-corrected paired t-tests revealed significant differences between all conditions.
Compared to Google Docs ({\small $m=1.9$}), participants identified significantly more problematic design choices when using both \ailegofull\ ({\small $m=5.6, p < .001$}) and \ailegolite\ ({\small $m=3.7, p = .03$}).
This indicates that \ailego\ equips cross-functional teams with the ability to explore and surface more potential harms during the AI development process.
Moreover, participants identified significantly more problematic design choices when using \ailegofull\ than \ailegolite\ ({\small $p = .04$}), suggesting that the additional \dvpersonacenteredeval\ feature further supports brainstorming by encouraging a more comprehensive exploration of possible design flaws.

\subsubsection{\ailego\ allows cross-functional teams to identify more likely and relevant AI system design flaws.}
The Friedman test revealed a significant main effect of condition on the likelihood of identified harms ({\small $\chi^2 = 9.72, p < .01$}). Post-hoc Holm-corrected Wilcoxon signed-rank tests showed that, compared to Google Docs ({\small $mdn=2$}), the harms identified using both \ailegofull\ ({\small $mdn=3.25, p=.03$}) and \ailegolite\ ({\small $mdn=3, p=.03$}) were rated as significantly more relevant. No significant difference was observed between \ailegolite\ and \ailegofull\ ({\small $p = .2$}).
This result underscores that \ailego\ enables teams to focus on more likely design flaws, improving their ability to prioritize relevant risks in the development process.
No significant main effect was found regarding the severity of identified harms, indicating that \ailegofull\ primarily enhances teams' capacity to detect risks that are more probable rather than more severe.

\subsubsection{\ailego\ tools are more preferred and show comparable usability.}
Although participants rated the highest average overall SUS score for \ailegolite\ in the planning phase and \ailegofull\ in the evaluation phase, statistical tests revealed no significant main effect, indicating the margins were negligible.
For individual SUS items in the planning phase, Holm-corrected Wilcoxon signed-rank tests showed that participants rated \ailegolite\ significantly lower on the item ``I found the tool very cumbersome to use'' ({\small $W = 0.00, p= .04$}) and significantly higher on the item ``I found the various functions in this system were well integrated'' ({\small $W = 0.00, p= .03$}) compared to Google Doc.
For the planning phase, 5 out of 6 participants preferred \ailego\ to Google Docs.
For the evaluation phase, the Friedman test showed a significant main effect on average ranking ({\small $\chi^2 = 10.67, p< .01$}).
Post-hoc Holm-corrected Wilcoxon signed-rank tests showed that participants ranked \ailegofull\ ({\small $mdn=1$}) significantly higher than Google Doc ({\small $mdn = 3, p< .001$}).

\subsection{Values Over Existing RAI Practices (RQ2)}
\label{sec:findings-values}
Interview notes from participants across teams (T1--6) with different roles (AI, UX, PM) revealed varied effectiveness in \ailego's features in achieving the design goals, while also identifying opportunities for future improvements.

\subsubsection{Eight-stage Worksheet \& Checklist serve as an effective channel for knowledge handoff in cross-functional teams (G1).}
Participants reported that they could more easily elaborate their ideas under the scaffolding provided by \dveightstageworksheet\ \& \textit{Checklist}. Both technical ($n=3$) and non-technical/user-facing roles ($n=7$) considered them being helpful in structuring their thoughts within the framework of the AI lifecycle. Compared to the ``free-form'' communication style of Google Docs (T2AI), this structured method helped participants focus on individual AI development stages, facilitating design choice elaboration and idea comprehension:

\begin{quote}
\textit{I can easily follow the prompts (from \dveightstageworksheet) to plan around the typical workflow of (AI system) development, whereas in Google Docs I have no clue of what to describe and where to start from.} (T5AI)

\textit{Following the clear structure by \ailego\ allows me to easily comprehend my teammate's planning.} (T6UX)
\end{quote}

Additionally, the eight-stage mechanism breaks down tasks for individuals, which makes communication more explicit and actionable, rather than being ambiguous and unapproachable (T3PM, T4AI, T4UX, T6UX):

\begin{quote}
\noindent\textit{\ailego\ provides better support in writing out the plan and specifies the tasks I should focus on more clearly.} (T4AI)
\end{quote}

Participants T1AI and T2AI explicitly reported feeling encouraged to think more broadly about critical components that they may otherwise have overlooked via Google Docs. They also described a shift towards more proactive risk identification from a reactive, ``fix-it-later'' approach (T2AI):

\begin{quote}
\textit{The ``Feedback'' stage in \ailego\ prompted me to consider input from a broader range of sources, including users, law enforcement, and other public entities.} (T1AI)
\end{quote}

The easy-to-grasp stage-based inputs provided by AI developers subsequently enabled non-technical/user-facing roles to engage more actively and take greater responsibility in the AI development process (T3PM, T4UX, T5UX), in contrast to traditional models where technical teams hold primary power in defining the project.
As T3PM noted, instead of being a ``passive recipient of information'', they became ``an active partner in the responsible AI process''.

\subsubsection{\ailego\ presents the workflow through an intuitive and engaging layout (G2).}
Most participants ($n=11$) found the tree taxonomy of interconnected \dvlifecycleblocks\ in \ailego\ suitable for the AI product design process.
Compared to plain text in Google Docs, the tree taxonomy allowed participants to view the entire flow of AI development at a glance. Instead of scrolling up and down to locate relevant notes, they could easily navigate the workflow and focus on specific stages. This design enabled participants to track their progress more effectively while maintaining a continuous thought process:

\begin{quote}
\textit{The tree taxonomy visualization allows me to keep track of my progress when drafting the plan. However, with Google Docs, I struggled to get an overview and frequently had to navigate back and forth to reference earlier designs.} (T3AI)
\end{quote}

This flow-based information architecture serves as a shared visual language, fostering a more organized and targeted approach to collaboration. Participants T1UX, T5UX and T6UX emphasized that it made them feel like integrated contributors to the development lifecycle rather than participants in a one-off feedback session:

\begin{quote}
\textit{The interaction mechanism of AI LEGO where you can hover on a block to see the prompt is very intuitive.} (T1UX)

\noindent\textit{I particularly like this kind of flow-based format...and it's a good communication model...for example, you can leave comments on a specific block, and then you can have an async conversation with the developer about a specific piece of content.} (T5UX)
\end{quote}

Building on this structure, \ailego\ further supports seamless user interaction. Participants praised \ailego\ for facilitating direct manipulation, a key concept around the sense of control in HCI, through integration between \dveightstageworksheet/\textit{Checklist} and specific UI features. For instance, participants appreciated that they could drag and drop stage blocks (T2UX) and hover over them to view prompts (T1UX, T4AI):

\begin{quote}
\textit{The interaction mechanism of AI LEGO where you can hover on a block to see the prompt is very intuitive.} (T1UX)

\textit{With the drag and drop feature (over the stage blocks), the experience is interactive and playful.} (T2UX)
\end{quote}

The intuitive and visually appealing overall layout led to a reduced learning curve and a sense of playfulness (T2UX, T4UX):
\begin{quote}
\textit{Even without an explanation of the persona-based evaluation feature, I can still learn how to use it through the provided prompts and self-exploration.} (T4UX)

\textit{\ailego\ gives a playful vibe in the theme of LEGO with the colorful blocks and minifigures.} (T4PM)
\end{quote}

\subsubsection{\ailego\ helps support the evaluation process by non-technical roles (G3).}
Compared to Google Docs, \ailego\ enabled non-technical roles to identify more potential harms of AI products ({\small $p<.001$}) and with a higher likelihood ({\small $p=.02$}).
When using Google Docs, non-technical participants frequently encountered inconsistent structure (T1PM, T3UX, T5UX) and unclear technical jargon (T1PM, T2PM, T3UX, T4UX, T5UX) in the drafted plans; in contrast, no such obfuscation was reported when using \ailego.
The integrated and scaffolded design of \ailego\ also reduces the cognitive load of switching between different information sources scattered across the document (T3UX, T5UX).
This design helps ``shift'' (T3UX) the focus of ethical analysis from abstract concepts to concrete, stage-specific problems, leading to a more deliberate and contextualized approach to harm analysis, which is more precise and actionable than the generalized, scattershot analysis often seen in Google Docs + Harm Envisioning (T1PM, T2PM).

Most of the non-technical participants ($n=8$) attributed the enhanced evaluation efficacy primarily to the \dvpersonacenteredeval. By engaging with the mocked personas, participants were able to ``rethink'' each stage of the AI development plan through the lens of identified stakeholders, allowing them to generate a broader range of ideas from multiple viewpoints ($n=6$) and directly link abstract technical plans to ``real-world human impact'' (T5PM, T4PM):
\begin{quote}
\textit{The generated personas allow me to think about a case more deeply based on their characteristics, which is really helpful in finding the edge cases.} (T1PM)

\textit{...the persona function helps me broaden my point of view by providing stakeholders' perspectives.} (T2PM)
\end{quote}

According to participants, this approach was particularly effective in unfamiliar scenarios (T3UX, T4UX) or when working independently without access to collaborative input (T3PM, T5UX, T6UX). For examples,
\begin{quote}
\textit{The persona feature is particularly helpful when I am unfamiliar with specific scenarios, as it allows me to approach the problem from different standpoints and levels of granularity.} (T3UX)

\textit{The persona feature helps me think about things like labor markets, financial impact...things that are not very obvious.} (T4UX)

\textit{Personas could stimulate your thinking process when you work alone by providing mocked data.} (T6UX)
\end{quote}

Interestingly, some participants in non-technical/user-facing roles mentioned a shift in power imbalance present in current cross-functional teams (identified in the formative study).
Participants described \dvpersonacenteredeval\ feature as a resourceful ``intelligent assistant'' (T5UX) or a ``creativity stimulant'' (T6PM) that simulates different stakeholder perspectives, which could give them a stronger set of arguments to use when questioning or challenging a developer's plan (T5UX, T6UX).
Meanwhile, \dveightstageworksheet\ facilitated clear articulation of AI design choices from AI developers, reducing non-technical roles' needs for frequent clarification from developers, enabling them to work more autonomously without worrying they were constantly bothering technical counterparts:
\begin{quote}

\textit{(Without \ailego) I would probably need to ask a few questions...things that I didn't understand, or things that they (AI developers) didn't explain clearly.} (T4UX)
\end{quote}

\subsubsection{Future improvements to better achieve the design goals.}
While participants generally appreciated \ailego\ for its enhanced efficiency and user experience, they also offered suggestions for improvements regarding the identified design goals.
To enhance the effectiveness of knowledge handoff (G1), participants proposed adding prompts tailored to specific AI products or scenarios in \dveightstageworksheet\ and \textit{Checklist} (T3AI) and enabling customizable blocks for additional notes.
T4UX suggested enhancing the intuitiveness of \ailego's visualization by incorporating formatted text similar to Google Docs (G2).
For \dvpersonacenteredeval\ (G3), participants recommended exploring features such as backend automation to trigger evaluations upon detecting issues (T3UX) and incorporating harm severity ratings to aid prioritization (T4UX). Participants T4UX and T6UX expressed concerns about the AI-simulated stakeholders' lack of real-world context or accuracy:

\begin{quote}
\textit{I'm a little concerned because it's all AI-generated...I don't know the truth or the priority (of the harms) without doing my own research.} (T4UX)
\end{quote}

Mechanisms should thus be developed to prevent AI teams from over-relying on \dvpersonacenteredeval\, replacing the need for user-facing professionals' expertise and research to validate and prioritize the identified harms. We further discuss this in Section~\ref{sec:discussion-hai}.
\section{Discussion and Design Implications}
Our findings highlight the effectiveness of \ailego\ in scaffolding cross-functional AI teams in handing off knowledge of high-level AI system designs and conducting more comprehensive early-stage harm evaluation.
Below, we explore the role of \ailego\ within three critical aspects of RAI, highlighting its contributions, limitations, and implications for fostering future RAI practices.

\label{sec:discussion-opportunities}
\subsection{Supporting Early-stage Cross-functional RAI Envisioning}

Our findings highlight early-stage RAI envisioning as a form of cross-functional coordination work rather than a purely technical or reflective activity. Compared to prior early-stage tools that primarily support individual practitioners or narrow design moments~\cite{wang2024farsight,lam2023model}, \ailego\ focuses on structuring how technical and non-technical roles collaboratively articulate, examine, and iterate on AI design decisions before implementation.
Concretely, \ailego\ introduces stage-based lifecycle blocks, associated prompts, and linked evaluation artifacts as shared representations through which teams coordinate this work.
In doing so, it foregrounds the collaborative labor involved in anticipating harms when system designs are still fluid and underspecified.

A key contribution of \ailego\ lies in its use of abstraction to make early AI design work collectively tractable. By breaking down AI development into modular, stage-based components, the \dveightstageworksheet\ and accompanying checklists enable participants to reason about complex systems at an appropriate level of granularity during planning. This aligns with prior work on sketching and abstraction in early design~\cite{yang2019sketching,yildirim2024sketching}, suggesting that early-stage representations need not be complete or precise to support meaningful collaboration. Instead, they must be interpretable and negotiable across roles

Participants’ responses further illustrate a central CSCW tension between structure and openness in collaborative work. While many participants valued the stage-based structure and tree-like organization of \ailego\ as a shared reference point, providing common vocabularies and locus for discussion, some (T2AI and T6PM) preferred the baseline’s free-form documentation for its flexibility and ability to support open-ended exploration. Rather than indicating a failure of structure, this tension echoes longstanding CSCW findings that coordination mechanisms must balance formalization with opportunities for situated interpretation. Our findings suggest that effective early-stage RAI tools should allow teams to fluidly move between structured articulation and open reflection, depending on the phase and purpose of collaboration.

These findings reinforce the role of boundary objects in early-stage RAI work. The lifecycle stages and modular blocks in \ailego\ function as boundary objects that are simultaneously stable enough to anchor discussion and flexible enough to accommodate diverse perspectives~\cite{star1989institutional,madaio2024tinker}. By externalizing high-level design intent into shared, manipulable artifacts, \ailego\ supports articulation work across roles and enables harm identification to be treated as an integral part of collaborative planning rather than a downstream audit task.
Future work should further explore how such boundary objects can be adapted to different organizational contexts and collaboration styles. In particular, there is an opportunity to investigate how personalization, partial structure, or role-specific views can coexist with shared grounding, enabling teams to tailor early-stage RAI envisioning without fragmenting collective understanding.

\subsection{Towards Human-AI complementarity in RAI Practices}
\label{sec:discussion-hai}
The use of LLM-generated personas in \ailego\ raises broader questions about how human and AI capabilities should be combined in ethical deliberation and harm identification. Traditionally, RAI evaluation has relied on human expertise through methods such as expert review, crowdsourcing, and structured audits~\cite{bigham2015human,madaio2020co,deng2023understanding}. Recent work has begun to explore how LLMs can assist in these processes by generating candidate harms or stakeholder perspectives~\cite{buccinca2023aha,wang2024farsight}.

Rather than delegating evaluative authority to LLMs, \ailego\ adopts a complementary design in which participants retain control over harm identification while using LLMs as a generative resource. In this configuration, persona-centered evaluation supports ideation and perspective-taking without replacing human judgment. This design aligns with the perspectives on human-AI collaboration that emphasize the importance of preserving human agency, accountability, and contextual reasoning when introducing AI assistance into collaborative work~\cite{fan2022human, holstein2023toward}.
Importantly, the LLM-generated personas in \ailego\ function not merely as informational outputs, but as boundary objects that provide a shared narrative reference that participants can collectively interpret, contest, and refine. For non-technical practitioners, these personas help surface edge cases and prompt reflection on unfamiliar technical decisions. For teams as a whole, they could offer a common frame for discussing potential harms across roles with asymmetric expertise.

At the same time, participants’ concerns about the realism and completeness of LLM-generated personas highlight a critical tradeoff. While such personas can increase the elasticity of boundary objects, enabling teams to rapidly explore a broader range of scenarios, they may also introduce synthetic abstraction that risks obscuring real-world complexity. Moreover, because these personas are generated based on generic language models (e.g., GPT-4) and limited contextual prompts, they may reproduce prevailing assumptions or biases present in the training data, and are unlikely to surface ``blind spots'' rooted in lived experience, cultural specificity, or structural marginalization. This tension reflects a broader challenge in machine-mediated collaboration: balancing efficiency and generativity with the need for grounded, accountable representations~\cite{madaio2024tinker}.

We view this tension not as a limitation of \ailego\ alone, but as an open design space for CSCW research. Future work should further investigate how human-AI complementarity can be operationalized in RAI practices, including when and how AI-generated artifacts should be surfaced, constrained, or contextualized to support ethical deliberation without displacing human responsibility~\cite{hemmer2021human}.

\subsection{Integrating into Industrial Cross-functional Product Workflows}
The scope of this work is to understand the primary challenges in conducting early-stage cross-functional RAI work, as well as the effectiveness of \ailego\ in addressing them, specifically, knowledge handoff by technical roles and harm envisioning by non-technical/user-facing roles.
Compared to other solutions that remain at a high level of representation (e.g., guidelines~\cite{microsoft_responsible_innovation_community_jury, ballard2019judgment}), \ailego\ was purposefully built as a collaborative web tool.
The embodiment of our tools aims to integrate a less-formalized socio-technical activity into early product design processes, blending product planning with harm anticipation without imposing excessive workloads.
The evaluation results demonstrated the feasibility of this concept in enhancing involvement across roles and improving overall evaluation performance.
Participants T2PM, T3PM and T4UX specifically praised how \ailego\ modularizes RAI work into stepwise tasks with clear instructions, noting its alignment with industrial practices. For example,
\begin{quote}
\textit{Given that industrial practices invest more in pushing the product, I think AI LEGO could work as it scales up the evaluation process through collaboration and AI generation without adding significant extra work.} (T4UX)
\end{quote}

While our system and study designs were largely grounded in realistic settings, we must acknowledge that industrial cross-functional practices involve additional factors and complexities that were not fully captured in this work~\cite{deng2023investigating,piorkowski2021ai}.
For example, different teams and organizations use varied communication channels, project planning tools, and collaboration policies that shape their working practices.
While \ailego\ can be further optimized for minimal onboarding effort or adapted into lightweight plugins or a fully featured tool for practical values (e.g., with work allocation and milestone tracking features), this would require a deeper understanding of organizational preferences, constraints, and potential frictions that may limit adoption. For instance, T6UX noted:

\begin{quote}
\textit{The tool...is to come up with potential issues or potentially, you know, places of bias. But that's also kind of coming at it from a, like, ``deficit mindset''. I hadn't even thought of, like, who in my organization would want something like this because of the way the tool was framed.} (T6UX)
\end{quote}

Additionally, \ailego\ focuses on facilitating knowledge articulation from technical AI developers to non-technical/user-facing roles, based on insights from our formative study that tasks related to AI system planning and harm evaluation should be delegated to developers and non-technical/user-facing roles, respectively.
However, the underlying assumption that these roles are always cooperative (e.g., developers readily integrating feedback into system designs) does not always hold.
Our qualitative results suggest that \ailego\ shows promise in addressing the prevailing power imbalance, enabling non-technical and user-facing roles to participate more autonomously and actively in responsible AI planning rather than relying on developers' voices. Nevertheless, more systematic investigations in future studies would be necessary to validate this, particularly through field deployments of \ailego\ in uncontrolled, real-world industrial team environments, to examine how \ailego\ may support, mitigate, or reshape broader organizational dynamics (e.g., power, accountability) as reflected in actual collective behaviors.

\subsection{Limitations and Future Work}
\label{sec:limitations}
As an initial effort to support early-stage, cross-functional RAI work, our approach has a number of limitations.
First, \ailego\ was developed primarily to address the challenge of knowledge handoff from technical to non-technical/user-facing roles in cross-functional RAI work---a challenge well-documented in prior literature and surfaced in our formative study as a common issue in industry (See Section 3).
Toward more human-centered RAI practices, we acknowledge that future work should also investigate alternative workflows---particularly knowledge flows from non-technical roles to technical ones. These workflows are critical for incorporating contextual, ethical, and user-centered insights into technical decision-making. 
Additional areas for future exploration include concept mapping~\cite{lam2023model}, as well as collective sensemaking and cross-role decision-making.

Second, as a proof-of-concept prototype, \ailego\ would benefit from further refinement to enhance user experience, performance, and workflow integration (see Section~\ref{sec:findings-values}).
In particular, as discussed in detail in Section~\ref{sec:discussion-opportunities}, the LLM-simulated \dvpersonacenteredeval\ holds both promises and risks and would require more systematic future investigation (e.g., higher versus lower levels of LLM autonomy in harm assessment or user inquiries). 
Please note that our LLM-based persona is intended only as a starting point---not a substitute for real-world stakeholder engagement~\cite{agnew2024illusion, fan2024user}. Future work should explore how to more effectively integrate real-world users and stakeholders into the pipeline we introduce here.  More nuanced data that reflect real end-user interactions and feedback is needed to provide stronger stakeholder-based evaluations in RAI practices. 

Third, while our evaluation study aimed to approximate real-world cross-functional collaboration settings, many simplifications were made in the study design for practical reasons.
For example, we only recruited participants based in the US and formulated teams, each consisting of three typical roles, which limited our understanding of how the effectiveness of our tool might vary across different cultures, regions, and team setups.
\ailego\ was evaluated end-to-end in a single pass, reflecting a minimal level of team coordination and offering limited opportunities for dynamic feedback and idea exchange. Future work should explore less controlled experimental settings, such as deployment studies, to more accurately mirror everyday practices.

Finally, the six AI product scenarios used in the evaluation were adapted from popular AI incidents. Although we counterbalanced their selection and order, we did not fully control for each scenario's complexity. Several participants noted that more complex scenarios appeared to heighten the tool's perceived value.
Future investigations could thus delve deeper into how scenario complexity influences the tool's effectiveness.

\section{Conclusion}
We introduce \ailego, an interactive tool designed to help cross-functional AI practitioners better communicate high-level AI designs throughout the development lifecycle and identify potential harmful design choices early on. We evaluated \ailego\ with 18 industry practitioners from various roles, including AI developers, UI/UX designers, and PMs. Our results show that \ailego\ helped participants identify more problematic design choices and those with higher likelihood.
This implies that the specific design of the tool would impact the effectiveness of facilitating cross-functional knowledge handoff and harm envisioning, such as the stage-level prompts, block-based layout, and LLM-simulated personas.
We discuss the challenges and opportunities for supporting cross-functional collaboration in early-stage RAI work.
\section*{Availability}
\ailego\ is live at \url{https://ailego.vercel.app/} (source code: \url{https://github.com/henrw/ai-lego}).
\bibliographystyle{ACM-Reference-Format}
\bibliography{references}

\appendix
\newpage

\section{Initial UI Designs in Formative Study}
\label{sec:codesign-ui}
\begin{figure}[H]
\centering
\includegraphics[width=0.95\textwidth]{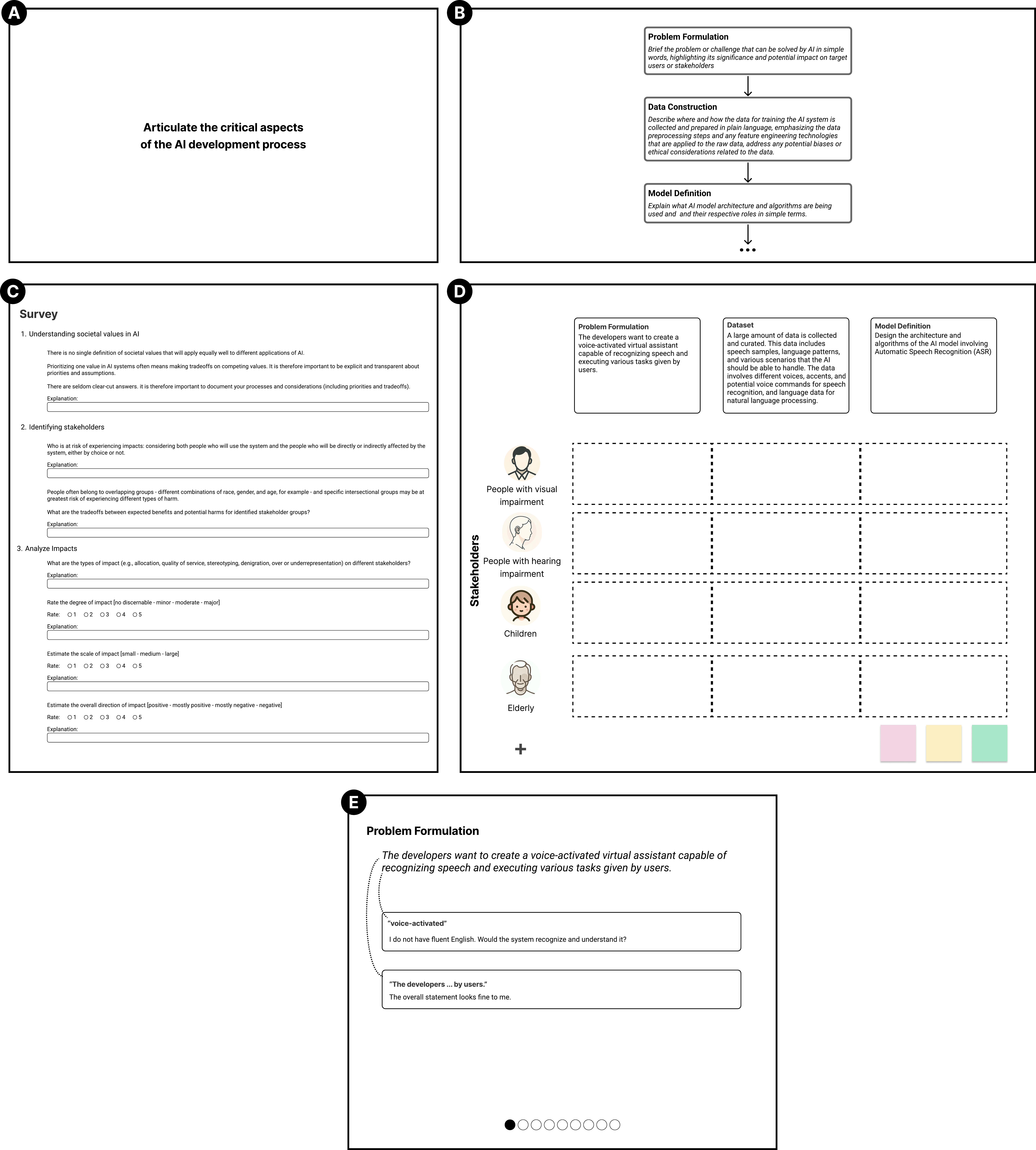}
\Description{Five prototype artifacts evaluated during the formative study. Two artifacts explore representations for AI development planning: an unstructured text description and a stage-based workflow. Three artifacts explore harm identification: a survey adapted from Value Cards, a stakeholder matrix relating AI development stages to stakeholder groups, and a commenting interface for discussion. These artifacts informed the design of AI LEGO.}
\caption{Concepts \& Artifacts explored in the formative semi-structured interview study. Planning: (A) Plain description presents currently unstructured approaches to articulating AI development plans in cross-functional teams; (B) Stage-based scaffolding breaks down and connects AI development stages while providing prompts, as inspired by AI storyboard~\cite{cramer2019challenges}. Harm identification: (C) The survey was adapted from Value Cards~\cite{shen2021value} investigating different social-technical dimensions, (D) Stakeholder table visualizes AI development stages versus stakeholders for better sensemaking processes, (E) Commenting represents common features in asynchronous communication tools.}
\label{fig:initial-uis}
\end{figure}

\newpage
\section{Formative Study Semi-Structured Interview Protocol}
\label{sec:formative-interview-protocol}

Note: Follow-up questions and probes were used throughout the sessions to clarify responses.

\subsection*{Background and Current Practice}
\textit{Goal: Understand participants’ roles, existing workflows, and baseline coordination practices.}

\begin{enumerate}
    \item \textbf{Role and context}
    \begin{itemize}
        \item Can you briefly describe your current role and responsibilities?
        \item Do you primarily work in a technical, non-technical, or mixed role when developing AI systems?
    \end{itemize}

    \item \textbf{Current tools and workflows}
    \begin{itemize}
        \item What tools do you currently use for project tracking, planning, or interdisciplinary collaboration (e.g., JIRA, Asana, Google Docs)?
        \item How do these tools support collaboration between technical and non-technical or user-facing roles?
    \end{itemize}

    \item \textbf{Demonstration and reflection}
    \begin{itemize}
        \item If possible, could you walk us through how you used one of these tools in a recent project?
        \item What aspects of the tool worked well? Where did you encounter friction or limitations?
        \item Please take a moment to write down any thoughts, preferences, or frustrations you have with these tools.
    \end{itemize}
\end{enumerate}

\subsection*{B. Experiences with Ethical Issues and Cross-functional Communication}
\textit{Goal: Surface concrete breakdowns in articulation, handoff, and harm-related reasoning.}

\begin{enumerate}
    \item \textbf{Concrete experience}
    \begin{itemize}
        \item Can you recall a specific instance where ethical or harm-related concerns arose during an AI project?
        \item How were these concerns identified, discussed, and addressed within your team?
    \end{itemize}

    \item \textbf{Communication challenges}
    \begin{itemize}
        \item Were there moments when communication between technical and non-technical roles was difficult? Why?
        \item What kinds of information were hardest to convey or understand across roles?
    \end{itemize}

    \item \textbf{Scenario-based probe (if no prior example)}
    \begin{itemize}
        \item Imagine your team is building an AI system for a given application scenario.
        \item How would your team plan the system and identify potential risks or harms?
        \item What challenges do you anticipate in this process?
    \end{itemize}
\end{enumerate}

\subsection*{AI Development Lifecycle and Planning Representations}
\textit{Goal: Probe reactions to different representations of early-stage AI planning.}

\begin{enumerate}
    \item \textbf{Lifecycle reflection}
    \begin{itemize}
        \item (Show AI development lifecycle figure~\cite{cramer2019challenges})
        \item Does this lifecycle resemble how AI development is planned in your work?
        \item Are there stages you would add, remove, or merge?
    \end{itemize}

    \item \textbf{Artifact comparison: planning}
    \begin{itemize}
        \item (Show Artifacts A and B)
        \item How effective is each approach for explaining key design choices?
        \item Which representation would better support collaboration across roles, and why?
        \item How might these representations affect early identification of potential issues?
    \end{itemize}
\end{enumerate}

\subsection*{D. Harm Identification and Evaluation Scaffolds}
\textit{Goal: Understand how participants reason about harms and evaluate different scaffolding approaches.}

\begin{enumerate}[nosep]
    \item \textbf{Artifact comparison: harm identification}
    \begin{itemize}
        \item (Show Artifacts C, D and E)
        \item How effective is each approach for harm identification?
        \item How would you use each of these in your current workflow?
        \item What do you find helpful or challenging about each approach?
    \end{itemize}
    \item \textbf{Comparative reflection}
    \begin{itemize}
        \item Which artifact(s) best support systematic evaluation of potential harms?
        \item Which better support open-ended discussion or iteration?
        \item How do these approaches differ in supporting technical versus non-technical roles?
    \end{itemize}
\end{enumerate}

\section{\ailego\ UI Components}
\label{sec:ui-components}
\begin{figure}[H]
\centering
\includegraphics[height=0.9\textheight]{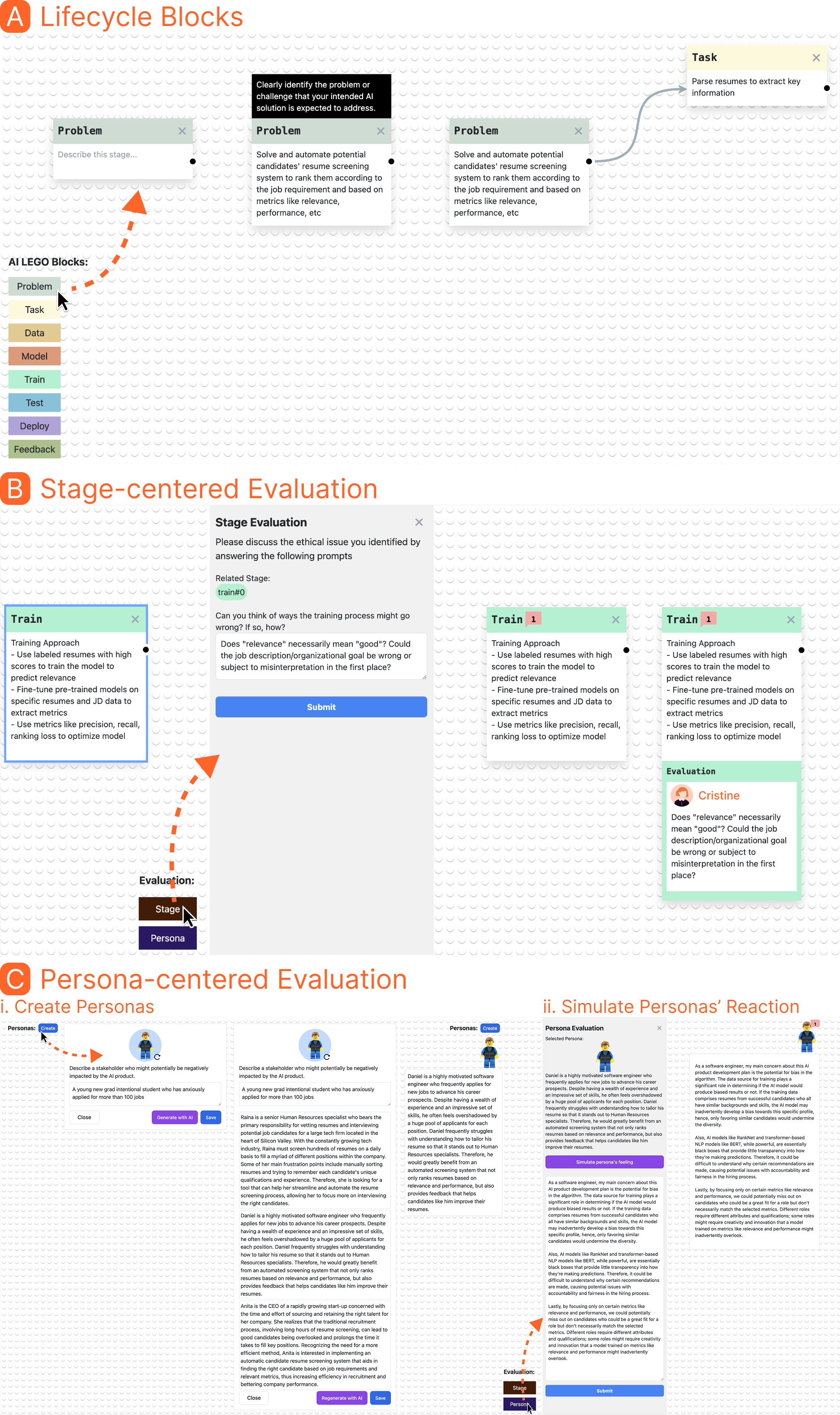}
\Description{Screenshots illustrating the major user interface components of AI LEGO. Lifecycle Blocks allow users to create connected AI development stages with guided prompts. Stage-centered Evaluation attaches structured evaluations to individual stages. Persona-centered Evaluation supports creating stakeholder personas and generating simulated stakeholder reactions to the proposed AI development plan.}
\caption{Key components and transitions in \ailego\ user interface.}
\label{fig:ui-components}
\end{figure}

\section{Prompts \& Examples for Persona-based Evaluation}
\label{sec:prompts}
\noindent\textbf{Persona Generation:}
\begin{framed}
\ttfamily\small
Generate three distinct personas, each described in a concise one-paragraph format. These personas should reflect individuals who are either directly or indirectly impacted by the identified problem(s): \{problem descriptions\}. Additionally, they should complement the Existing Personas by representing a broader spectrum of stakeholders. Each paragraph should begin with a brief summary of the persona, followed by a detailed account of their background, including relevant personal and professional aspects, motivations, and how they engage with the problem.
Existing Personas: \{personas\}
\end{framed}

\noindent \textit{Example Output:} Samantha is a Human Resource Manager at a fast-growing tech startup. In charge of recruitment and talent management, she has an average of 200 resumes to screen for each open position. Samantha has over a decade of experience in HR which has equipped her with a good eye for filtering resumes. But due to the surge in applications, her workload has significantly increased which is slowing down the hiring process considerably. The stress of potentially missing out on quality candidates due to a manual and time-consuming process affects her greatly. She is in search of an automated resume screening tool that can drastically reduce the time taken for this step and improve efficiency, allowing her to focus more on the subsequent interview and negotiation stages.\\

\noindent\textbf{Stakeholders' Feeling Simulation:}
\begin{framed}
\ttfamily\small
Assume you are \{persona description\}. You will be presented with a development plan for an AI product and asked to express your concerns about its potential negative impact on you. Your response should be clear, concise, and well-reasoned, aligning with your persona's background, lived experience, and perspective. 
AI Development Plan:\\
\{\{stage name\}: \{stage description\} for all stages\}
\end{framed}

\noindent \textit{Example Output:} While the proposed AI implementation certainly presents an innovative and efficient approach towards streamlining the hiring process, it raises certain concerns and implications from an HR perspective.

One of the primary concerns is accuracy. The AI performance would heavily depend on how well the algorithms are trained and fine-tuned. It might fail to detect subtlety, creativity, or potential in candidates' resumes that human judgment can ascertain. Therefore, we could lose out on good candidates because of the reductionist approach the AI might take to screen resumes. 

Secondly, the customization of models based on the resume-job fit is concerning. Given the dynamic nature of each job role and its requirements, the AI might not accommodate the nuances and specifics of each individual role.

Thirdly, bias is a significant concern while deploying AI models for recruitment. If the current employees' data is used for training these models, there's a risk of replicating biases that might exist in the present hiring process. The AI model could be biased against non-traditional career paths, gaps in employment, or non-conventional educational backgrounds. 

Finally, the ethical dimension of data privacy must be considered. Using public resume datasets or synthetic data shown during model training emphasizes the need for clear, transparent communication with applicants about how their personal information is being used.

\section{AI System Development Scenarios}
\label{sec:scenarios}
\subsection{Scenario 1: Image Generation}
\textbf{Background}: Creating custom visual content efficiently is essential for rapid prototyping, creative exploration, and personalized media production today. This service can also significantly boost a company's revenue.

\noindent\textbf{Project goal}: Develop an AI-powered image generation tool that creates high-quality, contextually relevant images from text prompts.

\noindent\textbf{Evaluation metrics}: Image quality; Generation speed; Model robustness.

\noindent\textbf{Key features}: Text-to-image conversion using advanced AI models; Customizable image style options.

\noindent\textbf{Requirements}: Utilize transformer-based models for image generation; Get the model trained from the selected dataset.

\noindent\textbf{Deployment}: Use cloud-based services for scalability and reliability. Regularly update the model with new data to improve quality.

\noindent\textbf{Target Users}: Graphic designers; Content creators; General consumers.
 
\subsection{Scenario 2: Gunshot Detection}
\textbf{Background}: Quicker and more accurate responses to gun-related incidents are crucial for enhancing urban public safety.

\noindent\textbf{Project goal}: Develop an AI-powered gunshot detection system that identifies and alerts gunshot sounds in real time.

\noindent\textbf{Evaluation metrics}: Detection accuracy; Response time; Reliability under different environments.

\noindent\textbf{Key features}: Real-time detection of gunshots; Automatic alerts to local police station; Integration with existing  infrastructure.

\noindent\textbf{Requirements}: Utilize a proper AI model for gunshot detection; Dataset needs to be created and marked by experts.

\noindent\textbf{Deployment}: Implement edge computing for real-time processing and scalability; Cloud-based monitoring. 

\noindent\textbf{Target Users}: Law enforcement agencies; Security Companies; Police.

\subsection{Scenario 3: Resume Screening}
\textbf{Background}: Manually processing a high volume of applications can be both time-consuming and resource-intensive in large organizations.

\noindent\textbf{Project goal}: Develop an AI-powered resume screening tool that automatically evaluates and ranks job candidates based on their resumes.

\noindent\textbf{Evaluation metrics}: Accuracy in extracting relevant information and ranking candidates.

\noindent\textbf{Key features}: Automatic extraction and analysis of key resume information.

\noindent\textbf{Requirements}: Use  NLP models for parsing and analyzing resume content. Datasets are given from internal sources from the company. Rank the candidates from multiple metrics.

\noindent\textbf{Deployment}: Integrate the AI tool with existing ATS or HR software platforms. Use cloud-based deployment for scalability and ease of updates

\noindent\textbf{Target Users}: HR Departments; Recruitment Agencies

\subsection{Scenario 4: Vaccine Allocation}
\textbf{Background}: Equitable and needs-based distribution of COVID-19 vaccines would maximize the safety of people.

\noindent\textbf{Project goal}: Develop an AI-powered vaccine allocation tool that prioritizes individuals based on urgency and need, ensuring equitable distribution during pandemics.

\noindent\textbf{Evaluation metrics}: Accuracy of risk stratification; Fairness in prioritization; Adaptability to new variants; Effectiveness of update.

\noindent\textbf{Key features}: Risk assessment based on health data, demographics, and exposure risk. Rank individuals by vaccination urgency; Real-time updates.

\noindent\textbf{Requirements}: Use  ML models for risk stratification and prioritization; Datasets are given from personal medical records and epidemiological data; Rank the candidates from multiple metrics.

\noindent\textbf{Deployment}: Integration with existing health information systems (EHRs, public health databases); Use cloud-based deployment for real-time updates.

\noindent\textbf{Target Users}: Public health organizations; Government agencies; Vaccine providers.

\subsection{Scenario 5: Credit Assessment}

\textbf{Background}: Traditionally, the process of determining creditworthiness has been complex and time-consuming, often resulting in lengthy waits for users.

\noindent\textbf{Project goal}: develop an AI-powered tool that evaluates and recommends credit card lines based on individual financial behavior and risk assessment.

\noindent\textbf{Evaluation metrics}: Accuracy compares to historical data (backtesting).

\noindent\textbf{Key features}: Assessment of creditworthiness using historical financial data; Risk scoring tool.

\noindent\textbf{Requirements}: Datasets are given from personal credit records and bank agent-approved data; Estimate the credit line based on multiple metrics from applicants' historical finance and credit data.

\noindent\textbf{Deployment}: Use secure cloud-based deployment for real-time updates; Integration with existing banking systems and customer profiles.

\noindent\textbf{Target Users}: Credit card issuers.

\section{Evaluation Study Semi-structured Interview Protocols}
\label{sec:eval-interview-protocol}
\noindent\textbf{Overall comparison.} Rank these tools from most to least effective for this task, and why? Were there moments when one tool made the task easier or harder?
    
\noindent\textbf{Understanding and handoff.} How easy was it to create (AI) or understand (UX, PM) the development plan in each condition? Did any tool reduce the need to ask for clarification?
    
\noindent\textbf{Workflow structure.} How did the organization and layout of each tool affect how you worked through the task?
    
\noindent\textbf{Harm identification (non-AI only).} Which tool best supported you in identifying potential harms? What features contributed most to this?
    
\noindent\textbf{Persona-centered evaluation (non-AI only).} If applicable, how did the persona feature influence your thinking? Were there cases where it felt unhelpful or unrealistic?
    
\noindent\textbf{Limitations and improvements.} What aspects of \ailego\ would you improve to better support RAI work in practice?

\newpage
\section{Evaluation Study Scenario \& Condition Orders}
\label{sec:assignment}
\begin{figure}[h]
    \centering
    \includegraphics[width=0.8\linewidth]{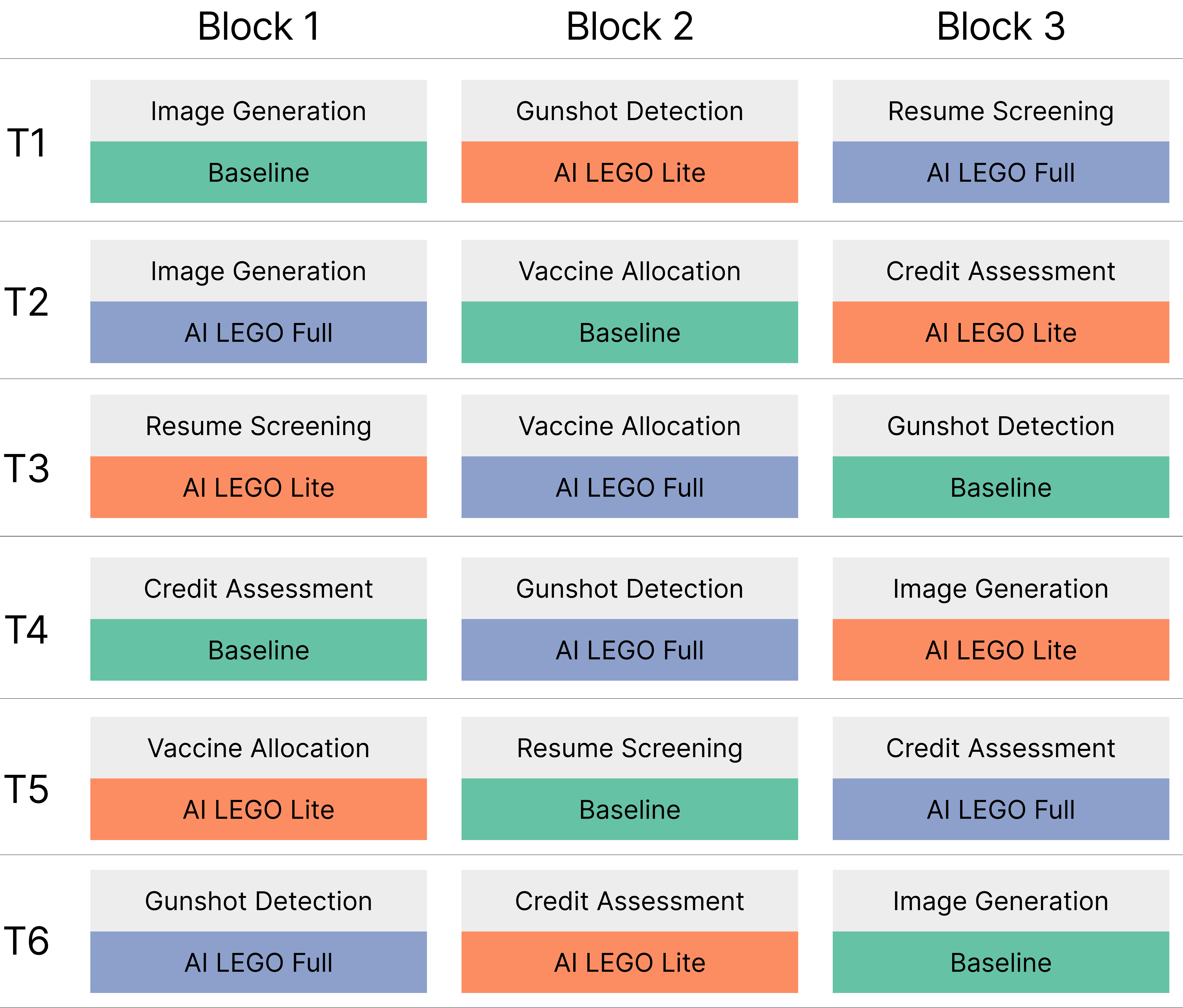}
    \Description{Matrix showing the assignment of five AI development scenarios and three experimental conditions across six study teams. Each team completed three blocks using Google Docs, AI LEGO Lite, and AI LEGO Full in counterbalanced order. Each scenario was evaluated under every condition multiple times.}
    \caption{Scenario \& Condition orders in the user study.}
    \label{fig:scenario-condition-order}
\end{figure}

\section{Harm Rating Rubrics}
\label{sec:rubric}

\begin{sideways}
\begin{minipage}{0.9\textheight}

\centering
\Description{Rubric used to rate identified harms during the evaluation study. The table organizes representative harms into four categories: representational harms, allocative harms, interpersonal harms, and social system harms. For each specific harm, it lists severity and likelihood rating ranges together with an example identified by study participants. The rubric was developed from the Sociotechnical Harm Taxonomy.}
\captionof{table}{Grading rubrics for harm ratings, developed using the Socio-technical Harm Taxonomy~\cite{shelby2023sociotechnical}. Specific ratings were determined based on the details and nuances of individual cases.}
\label{tab:rubric}

\resizebox{\linewidth}{!}{%

\begin{tabular}{p{37mm}p{50mm}p{30mm}p{30mm}p{70mm}}
\toprule
\textbf{Harm Type}            & \textbf{Specific Harm}                   & \textbf{Severity Range} & \textbf{Likelihood Range} & \textbf{Example} \\
\noalign{\global\arrayrulewidth=0.8pt}\midrule
\noalign{\global\arrayrulewidth=0.4pt}
\multirow{1}{*}{\textbf{Representational Harms}} 
                              & Unverified/Outdated/Inappropriate dataset                      & 3--4                   & 3--4 &  There could be biased data in the resume pool dataset for attributes like gender, races. The model trained on this will make biased decisions, which will reinforce people's biases. (G3PM)\\
                              & Uniformity of model metric               & 2--3                   & 2--3 & The choices of metrics would determine the social impacts (not hiring strong applicants for some other reasons; or hiring people who can only make money,) which might be unethical,l but fit well to the companies goal or social role. (G3UX)                   \\
\midrule
\multirow{1}{*}{\textbf{Allocative Harms}} 
                              & Inappropriate objectives in training leading to imbalanced inference        & 3--4                     & 3--4 & AUC score is not an appropriate metrics for "success" but instead should evaluate the model with human feedback. (G6UX)                      \\
\midrule
\multirow{1}{*}{\textbf{Interpersonal Harms}} 
                              & Privacy of service                      & 2--3                   & 3  &  (Gunshot detector) may cause privacy and social inequality for a certain neighborhood. (G4UX)                      \\
                              & Deploy data steal                       & 1                      & 4  & Using a cloud-based platform may lead to user data leaks. (G4UX)                      \\
                              & Inaccurate decision                     & 3--4                   & 2--4 & Innocent people (may be) identified as the shooter due to inaccurate gunshot detection (G3UX)                    \\
                              & Harm to specific users                  & 4                      & 4  & Image generator might not be able to generate images tailored to specific needs of customers (like logo designing for companies) if it uses a very general dataset of images. (G2UX)                     \\
                              & Data leak in training                   & 1                      & 3  & When training the model (for credit line assessment), there might be a data leak of personal information. (G2PM)                     \\
\midrule
\multirow{1}{*}{\textbf{Social System Harms}} 
                              & Explainability/Accountability           & 3--4                   & 3 & Model lack of explanation, too simple model(decision tree), may cause trustworthy issue for whoever using it. (G2PM)                       \\
                              & Copyright                               & 3                      & 3   & Data sources might have copyright issues for commercial uses that are stolen in the first place. (G2UX)                     \\
                              & Changed behavior                        & 1--4                   & 1--4 & Some users might change their financial behaviour based on the result of credit information gotten from AI tool. (G4PM)                    \\
                              & Insufficient feedback                   & 1--3                   & 1--3 & (The system should) have user input feedback rather than just thumb up and down. (G2PM)                    \\
\bottomrule
\end{tabular}
}
\end{minipage}

\end{sideways}
\end{document}